\documentclass[journal]{IEEEtran}
\usepackage[noadjust]{cite}
\usepackage{graphicx}
\usepackage[mathlines, switch]{lineno}
\usepackage{comment}
\usepackage[lofdepth,lotdepth, caption=false]{subfig}
\usepackage[x11names]{xcolor}
\usepackage{enumitem}
\usepackage{mdframed}
\usepackage{upgreek}

\usepackage{amsmath,amssymb}
\usepackage{color,soul}
\usepackage{calrsfs}
\usepackage[ruled]{algorithm2e}
\usepackage{setspace}
\makeatletter
\patchcmd{\@algocf@start}
  {-1.5em}
  {-5pt}
{}{}
\makeatother
\usepackage{multirow}
\usepackage{comment}
\newcommand{\squeezeup}{\vspace{-2.5mm}}
\usepackage{array}
\usepackage{makecell}
\usepackage{etoolbox}

\makeatletter
\patchcmd{\@makecaption}
  {\scshape}
  {}
  {}
  {}
\makeatother

\let\oldequation\equation
\let\oldendequation\endequation

\renewenvironment{equation}
  {\linenomathNonumbers\oldequation}
  {\oldendequation\endlinenomath}

\newcommand{\TA}[1]{\textcolor{black}{#1}}

\DeclareMathAlphabet{\pazocal}{OMS}{zplm}{m}{n}
\SetMathAlphabet\pazocal{bold}{OMS}{zplm}{bx}{n}

\newcommand*{\xhat}[1]{#1\kern-0.56em\hat{\phantom{#1}}}
\newcommand*{\xxhat}[1]{#1\kern -0.69em\hat{\phantom{#1}}}
\newcommand*{\xtilde}[1]{#1\kern-0.53em\tilde{\phantom{#1}}}
\DeclareMathOperator*{\argminA}{argmin}
\newcommand{\bbR}{\ensuremath{\mathbb{R}}}
\newcommand{\bbE}{\ensuremath{\mathbb{E}}}
\newcommand{\bbN}{\ensuremath{\mathbb{N}}}
\newcommand{\bbC}{\ensuremath{\mathbb{C}}}
\newcommand{\bbZ}{\ensuremath{\mathbb{Z}}}
\newcommand{\cW}{\ensuremath{\pazocal{W}}}

\newcommand{\cS}{\ensuremath{\pazocal{S}}}

\newcommand{\ba}{\ensuremath{\mathbf{a}}}

\newcommand{\bg}{\ensuremath{\mathbf{g}}}

\newcommand{\bh}{\ensuremath{\mathbf{h}}}

\newcommand{\bq}{\ensuremath{\mathbf{q}}}
\newcommand{\bs}{\ensuremath{\mathbf{s}}}
\newcommand{\by}{\ensuremath{\mathbf{y}}}

\newcommand{\bc}{\ensuremath{\mathbf{c}}}

\newcommand{\bp}{\ensuremath{\mathbf{p}}}
\newcommand{\bd}{\ensuremath{\mathbf{d}}}

\newcommand{\balpha}{\ensuremath{\boldsymbol{\alpha}}}

\newcommand{\btau}{\ensuremath{\boldsymbol{\tau}}}
\newcommand{\bmu}{\ensuremath{\boldsymbol{\mu}}}
\newcommand{\bphi}{\ensuremath{\boldsymbol{\phi}}}

\newcommand{\bPsi}{\ensuremath{\mathbf{\Psi}}}
\newcommand{\bUpsilon}{\ensuremath{\mathbf{\Upsilon}}}

\newcommand{\bPhi}{\ensuremath{\mathbf{\Phi}}}
\newcommand{\bTheta}{\ensuremath{\mathbf{\Theta}}}
\newcommand{\bPi}{\ensuremath{\mathbf{\Pi}}}
\newcommand{\bW}{\ensuremath{\mathbf{W}}}
\newcommand{\bX}{\ensuremath{\mathbf{X}}}

\newcommand{\bM}{\ensuremath{\mathbf{M}}}
\newcommand{\bP}{\ensuremath{\mathbf{P}}}
\newcommand{\bH}{\ensuremath{\mathbf{H}}}
\newcommand{\bF}{\ensuremath{\mathbf{F}}}
\newcommand{\bC}{\ensuremath{\mathbf{C}}}

\newcommand{\bA}{\ensuremath{\mathbf{A}}}
\newcommand{\bR}{\ensuremath{\mathbf{R}}}

\newcommand{\bD}{\ensuremath{\mathbf{D}}}

\newcommand{\bI}{\ensuremath{\mathbf{I}}}

\newcommand{\bT}{\ensuremath{\mathbf{T}}}
\newcommand{\bU}{\ensuremath{\mathbf{U}}}

\newcommand{\bQ}{\ensuremath{\mathbf{Q}}}

\newcommand{\bzero}{\ensuremath{\mathbf{0}}}
\newcommand{\diag}{\ensuremath{\text{diag}}}
\newcommand{\tr}{\ensuremath{\text{Tr}}}
\newcommand{\bcH}{\ensuremath{\boldsymbol{\pazocal{H}}}}

\newcommand{\bcX}{\ensuremath{\boldsymbol{\pazocal{X}}}}
\newcommand{\bcQ}{\ensuremath{\boldsymbol{\pazocal{Q}}}}

\newcommand{\bLam}{\ensuremath{\boldsymbol{\Lambda}}}
\newcommand{\larrow}{\leftarrow}

\SetKwInput{KwInput}{Input}
\SetKwInput{KwOutput}{Output}
\SetKwFunction{FHank}{hankel}
\SetKwFunction{FCons}{construct}
\SetKwFunction{FEsimateWeig}{estWeight}
\SetKwFunction{FWeigSubFit}{solNLS}
\SetKwFunction{FLS}{solLS}

\SetKwFunction{SVD}{SVD}
\SetKwFunction{TSVD}{TSVD}
\SetKwFunction{rowstack}{rowstack}
\SetKwFunction{colstack}{colstack}

\begin{document}
\bstctlcite{IEEEexample:BSTcontrol}

\title{Delay Estimation for Ranging and Localization Using Multiband Channel State Information
\thanks{The authors are with the faculty of Electrical Engineering, Mathematics
and Computer Science, Delft University of Technology, 2826 CD Delft, The
Netherlands. Dr.\ J.\ Romme is also with IMEC Holst Centre, Eindhoven, The Netherlands. E-mails:
\{t.kazaz, g.j.m.janssen, j.p.a.romme, a.j.vanderveen\}@tudelft.nl.
This research was supported in part by NWO-STW under contract 13970 (“SuperGPS”). A part of this work was presented at the Asilomar Conference on Signals, Systems, and Computers, Nov. 2019 \cite{kazaz2019time}. This is the author (before review) version of the work that is accepted for publishing in IEEE Transaction on Wireless Communications. For the final version of the work, please check IEEE Xplore.
}}%
%
%
%

\author{Tarik~Kazaz,~\IEEEmembership{Student~Member,~IEEE,}
        Gerard~J.~M.~Janssen,
        Jac~Romme,
        and~Alle-Jan~van~der~Veen,~\IEEEmembership{Fellow,~IEEE}}


\maketitle
\squeezeup
\squeezeup
\squeezeup
\squeezeup
\squeezeup
\squeezeup
\begin{abstract}
    In wireless networks, an essential step for precise range-based localization is the high-resolution estimation of multipath channel delays. The resolution of traditional delay estimation algorithms is inversely proportional to the bandwidth of the training signals used for channel probing. Considering that typical training signals have limited bandwidth, delay estimation using these algorithms often leads to poor localization performance. To mitigate these constraints, we exploit the multiband and carrier frequency switching capabilities of wireless transceivers and propose to acquire channel state information (CSI) in multiple bands spread over a large frequency aperture. The data model of the acquired measurements has a multiple shift-invariance structure, and we use this property to develop a high-resolution delay estimation algorithm. We derive the Cram\'er-Rao Bound (CRB) for the data model and perform numerical simulations of the algorithm using system parameters of the emerging IEEE 802.11be standard. Simulations show that the algorithm is asymptotically efficient and converges to the CRB. To validate modeling assumptions, we test the algorithm using channel measurements acquired in real indoor scenarios. From these results, it is seen that delays (ranges) estimated from a multiband CSI with a total bandwidth of 320 MHz show an average RMSE of less than $0.3$ ns ($10$ cm) in $90$\% of the cases.
\end{abstract}

\begin{IEEEkeywords}
Delay estimation, ranging, super resolution, subspace fitting, multiband CSI, IEEE 802.11be, WiFi-7, OFDM.
\end{IEEEkeywords}

\IEEEpeerreviewmaketitle

\squeezeup \squeezeup
\section{Introduction}

\IEEEPARstart{L}{ocation} awareness is of great interest in different areas related to navigation and sensing \cite{win2018efficient}, and it fosters a wide range of emerging applications such as crowd sensing \cite{capponi2019survey}, autonomous driving \cite{8246850} and assisted living \cite{witrisal2016high}.  These applications demand omnipresent and decimeter-level accurate localization. Traditionally, the Global Positioning System (GPS) is used at the core of almost all navigation systems. However, GPS signals are severely attenuated and impaired by multipath propagation effects present in harsh radio environments, such as indoor or urban canyons, resulting in poor localization \cite{kassas2017hear}.  Unfortunately, these environments are the ones where precise localization is needed the most.

\begin{figure*}[t]
	\centering
	 \includegraphics[width=14.5cm]{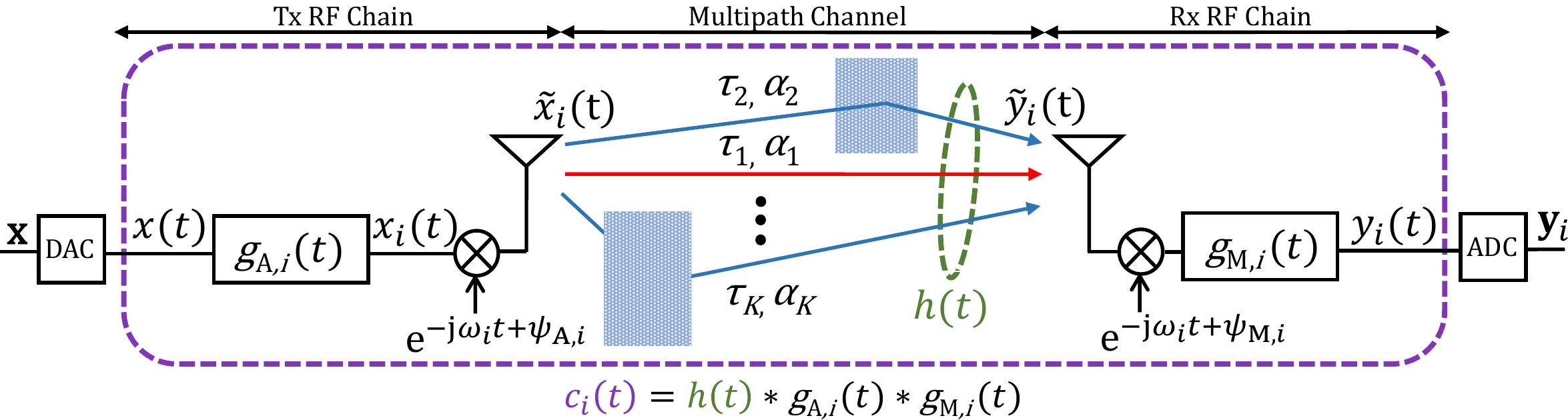}%
	 \caption{The multipath propagation environment between a mobile node and an anchor, described by $K$ MPCs, where red color denotes the LOS path. Each MPC is characterized by its complex gain $\alpha_k$ and delay $\tau_k$. The equivalent baseband channel (including the effects of the RF chains) is $c_i(t)$, where index $i$ refers to the $i$th band in a multiband system.}
	\label{fig:ilu1}
\end{figure*}

A promising localization approach in GPS-denied environments is to utilize existing wireless infrastructure and ambient radio frequency (RF) signals \cite{kassas2017hear}. Localization using these signals starts with the estimation of the multipath channels between the mobile node and multiple anchors \cite{leitinger2015evaluation}. Each channel is modeled as a sum of multipath components (MPCs), parametrized by their complex amplitudes, directions-of-arrival (DOAs) and delays, as shown in Fig. \ref{fig:ilu1}. In particular, the delay of the line-of-sight (LOS) path is directly linked to the range (distance) of the mobile node to the anchor, and forms the input for range-based localization methods based on time-of-arrival (TOA) and time-difference-of-arrival (TDOA). The localization performance of these methods primarily depends on channel estimation and the ability to resolve MPCs, estimate their parameters, and detect the LOS path.

Classically, delay estimation is based on searching the first dominant peak in the correlation between the received signal and the known training signal \cite{1458289}. The resolution of such methods is limited by the inverse of the bandwidth of the training signal. Typical training signals used in wireless networks have a fairly low bandwidth due to RF spectrum regulations and hardware constraints. The insufficient resolution prevents the separation of the LOS path from closely arriving MPCs, leading to biased range estimates and degraded localization performance \cite{wymeersch2012machine}. Therefore, the main challenge is the design of \textit{(i)} a practical approach for measuring the channel, and  \textit{(ii)} high-resolution delay estimation algorithms in the presence of close-in multipath.

\squeezeup

\subsection{Related Works}
Channel estimation is fundamental to wireless communications, but many works use this information for equalization \cite{bajwa2010compressed, meng2011compressive, chu2018semi}, where the precise knowledge of multipath parameters is not crucial. \TA{Methods for high-resolution channel estimation typically formulate the problem of delay estimation in the frequency domain, where it becomes a problem of parameter estimation of superimposed complex exponentials. The classical approaches to estimate these parameters are based on maximum likelihood (ML) estimation methods. Depending on the statistical assumptions made on the parameters, these methods can be classified into deterministic ML (DML) \cite{dml1} or stochastic ML (SML) \cite{sml1}. The SML estimators are asymptotically consistent and statistically efficient when the number of measurements increases to infinity \cite{kay1998estimation}. However, these methods involve solving the complex task of minimizing a nonconvex objective function with a highly multimodal shape, and many local minima \cite{ziskind1988maximum}. This optimization problem is typically solved using iterative algorithms \cite{yuan2011recent}, which require accurate initialization and, at best, guarantee convergence to a local minimum \cite{badiu2017variational}. Therefore, the performance of these methods depends highly on the accuracy of the initialization.}  Other methods for high-resolution delay estimation exploit the sparse nature of MPCs, and can be classified into those based on \textit{(i) subspace estimation} \cite{vanderveen1998estimation, li2004super}, \textit{(ii) compressive sensing} (CS) \cite{baraniuk2007compressive}, and \textit{(iii) finite-rate-of-innovation} (FRI) sampling \cite{1003065, gedalyahu2011multichannel, barbotin2012estimation}.  

\TA{When formulated in the frequency domain, the problem of delay estimation is a classical array signal processing problem, and methods such as MUSIC \cite{li2004super}, ESPRIT \cite{vanderveen1998estimation}, and Matrix Pencil \cite{hua1992estimating} are applicable. Moreover, when measurements are collected using the antenna arrays, these methods can be extended to two-dimensional (2-D) methods for joint angle and delay estimation \cite{hua1992estimating, van1998joint, gaber2014study}. Our work follows a similar approach to those in \cite{hua1992estimating} and \cite{van1998joint}, and we also propose subspace based method that exploits the structure presented in the measurements to estimate parameters of the channel model. However, the multiband channel measurements that we consider in this work have a 1-D model with a multiple shift-invariance structure, and we focus on designing an algorithm that will exploit this structure to increase the resolution of delay estimation.}

Compressed sensing (CS) methods exploit the sparse structure of multipath channels \cite{bajwa2010compressed}, in particular in UWB \cite{paredes2007ultra} and OFDM systems \cite{meng2011compressive}. Grid-based methods confine the MPC delays to a discrete set of predefined values. This causes basis-mismatch effects that are limiting resolution and leading to biased estimation with these methods. \TA{The problem of basis mismatch is solved using gridless sparse estimation algorithms \cite{chu2018semi}. Most of these methods transform the problem of frequency (i.e., delay) estimation into the estimation of a Toeplitz covariance matrix with low rank and positive semidefinite properties. Once the covariance matrix is estimated, the frequencies can be retrieved from its Vandermonde decomposition.  However, for the multiband channel state information (CSI), the covariance matrix will have a Toeplitz structure only when the measurements are collected in consecutive frequency bands. Therefore many of these methods can not be used for the estimation of the general multiband CSI models. The Bayesian view on the problem of gridless sparse estimation of complex exponentials is taken in \cite{badiu2017variational, hansen2018superfast}. In these works, the stochastic ML model regularized by sparsity promoting prior on the coefficients of the exponentials is used to describe measurements. These algorithms, in general, have high estimation accuracy and inherently estimate the number of MPCs present in the channel. In particular, the VALSE algorithm allows gridless estimation of complex exponentials with automatic estimation of the number of MPCs from incomplete, but single snapshot measurements \cite{badiu2017variational}. However, this is an iterative algorithm and has high computational complexity due to the variational estimation of the posterior frequencies. Its per-iteration complexity is cubic in the number of exponentials, and therefore, its complexity increases rapidly with the number of MPCs.}
 
\TA{Interesting results related to multipath channel estimation using the finite rate of innovation (FRI) framework are presented in \cite{gedalyahu2011multichannel}. This work shows that multichannel sampling with frequency mixing, i.e., multiband sampling, offers additional degrees of freedom that can further increase the resolution of delay estimation. However, in the proposed sampling method, the number of channels is proportional to the number of MPCs, making the proposed method impractical for wireless systems.}

The practical route to improved delay resolution is based on multiband channel probing \cite{xiong2015tonetrack,vasisht2016decimeter,chen2016achievingFH,khalilsarai2019wifi}. Here, multiple frequency bands are used to increase the frequency aperture of CSI measurements. Calibration is needed to undo the effects of transceiver impairments such as frequency and phase offsets that affect each band differently \cite{tadayon2019decimeter}. In \cite{xiong2015tonetrack}, MUSIC is used for delay estimation. However, this approach does not exploit all structures present in the multiband CSI, which results in statistically inefficient estimation.  In \cite{vasisht2016decimeter,khalilsarai2019wifi}, compressed sensing algorithms based on $\ell1$-norm regularized least squares (CS(L1)) are proposed. However, these algorithms consider the collection of CSI in consecutive bands and have limited resolution due to basis mismatch.  \TA{For additional comparisons in Section \ref{sc:num_exp}, we simulate performance of DML \cite{dml1} and SML \cite{sml1} algorithms and DOA estimation algorithms ESPRIT \cite{roy1989esprit}, and MI-MUSIC  \cite{swindlehurst2001exploiting} tailored to the problem of multiband delay estimation.}
\squeezeup

\subsection{Contributions}
\label{sc:contributions}
\begin{figure*} 
\parbox[t]{.49\textwidth}{ 
    \includegraphics[width=0.90\columnwidth]{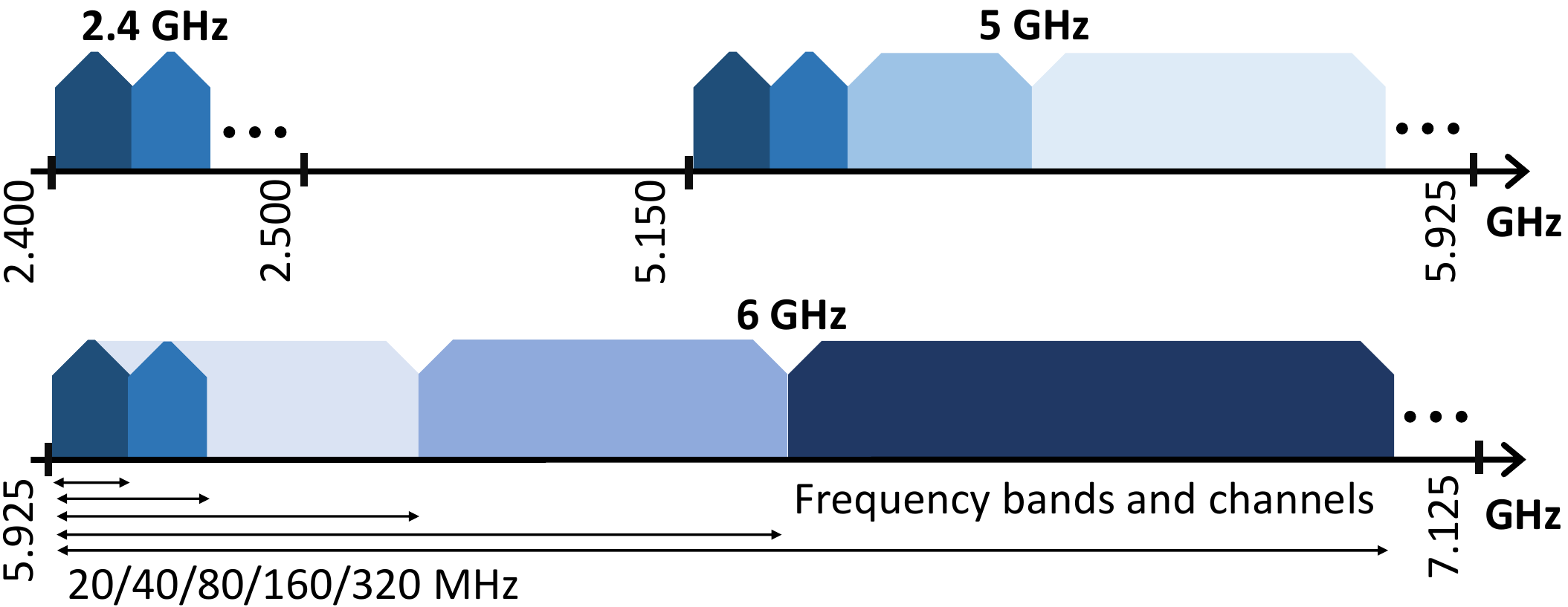}
    \caption{Example multiband system: Frequency bands defined for use in the IEEE 802.11be standard at  2.4, 5, and 6 GHz, with bandwidths of 20, 40, 80, and 160 MHz.}\label{fig:fig_ilu3}
}
\hfill
\parbox[t]{.49\textwidth}{
	 \includegraphics[trim=1 1 0 1,clip, width=9cm]{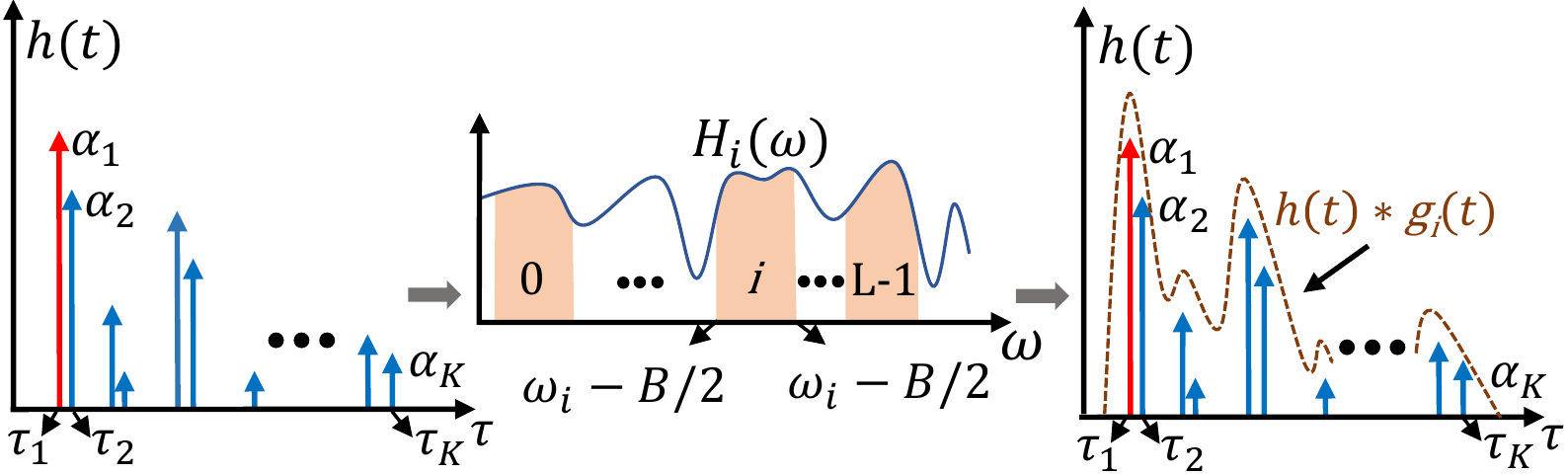}%
     \caption{The multiband channel probing and effects of the limited bandwidth of the transceivers on the delay resolution of multipath components.}\label{fig:ilu2}
}
\end{figure*}

In this paper, we exploit the multiband and carrier frequency switching capabilities of modern wireless transceivers and propose to acquire the CSI on multiple bands spread over a large frequency aperture to increase delay resolution. We start by deriving the data model for multiband CSI, considering orthogonal frequency-division multiplexing (OFDM) training signals as used in WiFi networks. \TA{The first difference of our work compared to the state-of-the-art methods presented in \cite{xiong2015tonetrack,vasisht2016decimeter,khalilsarai2019wifi} is the observation that by stacking the multiband CSI into Hankel matrices, the data model shows a multiple shift-invariance structure known from DOA estimation problems \cite{swindlehurst1992multiple, viberg1991sensor}. We use these properties to develop an algorithm that supports delay estimation from multiband CSI collected in arbitrary frequency bands. This is new compare to the algorithms proposed in \cite{xiong2015tonetrack,vasisht2016decimeter,khalilsarai2019wifi} which are restricted to delay estimation from the CSI collected in the consecutive bands. Furthermore, the proposed algorithm is gridless and does not suffer from basis mismatch, which is different than algorithms presented in \cite{vasisht2016decimeter, khalilsarai2019wifi}.  In our initial work \cite{kazaz2019time}, we considered this model and proposed a basic multiband delay estimation algorithm without weighting and data extension techniques. In the present paper, we extend on this and make the following additional contributions.}
\begin{itemize}[label={--}]
    \item We propose a weighted subspace fitting algorithm for delay estimation to exploit the multiple shift-invariance structure present in the multiband CSI.  We present the optimal weighting and introduce several data extension techniques that further improve the performance of the algorithm. After delay estimation, the complex amplitudes of MPCs are estimated by solving a linear least-squares (LS) problem.

    \item  We derive the Cram\'er-Rao Bound (CRB) for the multiband CSI data model, and analyze the effects of wireless system parameters, e.g., bandwidth, number of CSI measurements, and band selection, on the CRB.
    
    \item  We demonstrate the applicability of the proposed algorithm to the problem of delay estimation in the future WiFi-7 network defined by the IEEE 802.11be standard \cite{lopez2019ieee}. This standard will support multiband operation in 2.4, 5, and 6 GHz bands (cf.\ Fig.\ \ref{fig:fig_ilu3}), which makes this application an interesting showcase for the proposed algorithm. Various scenarios are simulated to show the influence of wireless system parameters on the algorithm's performance. 
    
    \item Motivated by experiments with a real indoor multipath channel dataset, we discuss the problem of band selection, considering the frequency dependency effects of RF signal scattering. These effects have been ignored in previous related works. However, they introduce modeling errors and deteriorate the performance of estimation if the frequency aperture is too large compared to the central frequencies of the bands. 
\end{itemize}

\squeezeup
\subsection{Organization of the Paper and Notation}
This paper is organized as follows. The system and data model are described in Section \ref{sc:system_model}. Section \ref{sc:algorithm} contains a detailed derivation of the basic steps of the algorithm, including several data extension techniques to improve robustness to noise and increase delay resolution. The Cram\'er-Rao Bound (CRB) is derived in Section \ref{sc:crb}. In Section \ref{sc:num_exp}, numerical simulations are performed to benchmark the algorithm's performance. Experiments with real channel measurements are elaborated in Section \ref{sc:real_exp}. Finally, conclusions are provided in Section \ref{sc:conclusions}.

The notation used in this paper is as follows. Bold upper (lower)-case letters are used to define matrices (column vectors). \TA{In particular, the bold letters indexed by subscript $i$ and superscript $m$ denote vector or matrices corresponding to $i$th band and $m$th snapshot, respectively. Otherwise, the letters without subscripts and superscripts denote vectors or matrices corresponding to overall multiband data.} $\bI_{N}$ and $\bzero_{N}$ denote $N \times N$ identity matrix and zero vector of size $N$, respectively. $\diag(\cdot)$ constructs a diagonal matrix from its vector argument.  \TA{$(\cdot)^T$, $(\cdot)^{*}$, $(\cdot)^H$, $(\cdot)^{\dagger}$, and $(\cdot)^{-1}$ represent transpose, complex conjugate, complex conjugate transpose, pseudo-inverse, and inverse of a matrix, respectively.} $\tr(\cdot)$ denotes the matrix trace operator and $\left\| \cdot \right\|_{F}$ is the Frobenius norm of a matrix. The Kronecker and Hadamard products are denoted with $\otimes$ and $\odot$, respectively. $\pazocal{CN}(\bmu, \sigma^2 \bI_N)$ represents a complex Gaussian normal distribution with mean $\bmu$ and covariance matrix $\sigma^2 \bI_N$.

\section{Data Model}
\label{sc:system_model}
In this section, we introduce the communication scenario, and derive the corresponding data model. To be relevant to current WiFi standards, we consider a localization system that uses OFDM training signals exchanged at multiple bands, i.e., frequency channels, to obtain multiband CSI measurements. We will refer to WiFi frequency channels as frequency bands. We first define continuous-time signal models for training signals and the multipath channel. We then derive the data model for multiband CSI, which reveals the multiple-shift invariance structure of the measurements. Finally, we briefly discuss synchronization impairments between transceivers and the impact of phase offset on the measurements.

\squeezeup
\squeezeup
\subsection{System Model}
Consider a localization system that uses OFDM training signals to estimate ranges between the mobile node and at least three (four) anchors for localization in 2-D (3-D) space. This process starts with the exchange of a known training signal $x(t)$ between an anchor and the mobile node (or vice versa), and estimation of the corresponding multipath channel (cf.\ Fig.\ \ref{fig:ilu1}). Assume that the training signal has $N$ orthogonal subcarriers in a single OFDM symbol where the symbol duration is $T_{\textrm{sym}}$ and the frequency spacing of adjacent subcarriers is $\omega_{\textrm{sc}}=2\pi/T_{\textrm{sym}}$. The duration of each OFDM symbol is periodically extended with the cyclic prefix of duration $T_{\textrm{cp}}$ to ensure cyclic convolution with the channel, which results in the total duration of a transmission block $T = T_{\textrm{sym}} + T_{\textrm{cp}}$. The baseband model for the training signal in a single transmission block can be written as
\begin{equation}
    \label{eq:pilot_sig}
       x(t)  = \left[\sum_{n=0}^{N-1} s[n] e^{j \omega_{\textrm{sc}}nt}\right] 
               p(t - T)\,,
\end{equation}
\noindent where $\bs = \left[s[0],\dots, s[N-1]\right]^T \in \bbC^N$ are the known training symbols, and 
\begin{equation}
    \nonumber
    \label{eq:trans_block}
       p(t)  =
       \begin{cases}
       \mbox{1}, & t \in [-T_{\textrm{cp}}, T_{\textrm{sym}}]\,,
       \\
       \mbox{0}, & \text{otherwise}\,.
       \end{cases}
\end{equation}
\noindent This training signal is upconverted to the carrier frequency $\omega_i$ and transmitted as 
\begin{equation}
    \label{eq:passband}
       \tilde{x}_i(t) = \text{Re} \left\{x(t) e^{j (\omega_i t + \psi_{\textrm A, i})} \right\} \,,
\end{equation}
\noindent where $\psi_{\textrm A, i}$ is an unknown phase of the local oscillator at the anchor node. Here we assumed without loss of generality that all frequency channels use the same training signal, as is often done in practice.
We further consider that the anchor and mobile node are frequency synchronized during channel probing. This is typically the case in practical OFDM systems, where before channel estimation, a frequency offset is estimated and compensated using known training signals such as legacy short and long training fields (L-STF and L-LTF) in IEEE 802.11be \cite{lopez2019ieee}.

\squeezeup
\subsection{Multiband Channel Probing}
To probe the multipath channel, the training signal $x(t)$ is transmitted at $L$ separate frequency bands, ${\cW_i = [\omega_i-\frac{B}{2}, \omega_i + \frac{B}{2}]}$, where $B$ is the bandwidth and $\omega_i$ is the central angular frequency of the $i$th band (cf.\ Fig.\ \ref{fig:ilu2}). 
We consider that the multipath channel is probed over a large frequency aperture. Therefore, it is suitable to use the UWB channel model \cite{molisch2009ultra} to model the propagation between the anchor and the mobile node. For this channel, the continuous-time channel impulse response (CIR) $h(t)$ is described as
\begin{equation}
\label{eq:chan_imp}
	h(t) = \sum_{k=1}^{K} \alpha_k \delta(t-\tau_k)\,.
\end{equation}
In this model, there are $K$ resolvable MPCs, where the $k$th MPC is characterized by its time-delay $\tau_k \in \bbR_+$ and its complex path amplitude $\alpha_k \in \bbC$. The time-delays are sorted in increasing order, i.e., $\tau_{k-1} < \tau_{k}$, ${k = 2, \dots, K}$, and $\tau_1$ is considered to be the LOS path. The complex path amplitudes ${\alpha_k \sim \pazocal{CN}(0, \sigma_{\alpha,k}^2)}$, ${k=1,\dots, K}$, have the average power $\sigma_{\alpha,k}^2$, and are assumed to be wide-sense stationary and mutually uncorrelated. 

Practical wideband antennas and RF chains have a frequency-dependent response \cite{gentile2020methodology}. We model the compound frequency response of the RF chains including antennas at the $i$th probed band as an equivalent linear and time-invariant baseband filter with impulse response $g_i(t) = g_{\textrm A, i}(t) \ast g_{\textrm M, i}(t)$ (cf.\ Fig.\ \ref{fig:ilu1}). Here, $g_{\textrm A, i}(t)$ and $g_{\textrm M, i}(t)$ are the impulse responses of the RF chains at the transmitter and receiver, respectively. The filter $g_i(t)$ has frequency response $G_i(\omega)$ with passband $\omega \in [-\frac{B}{2}, \frac{B}{2}]$. Then, the compound impulse response of the multipath channel and RF chains at the $i$th band is given by
\begin{equation}
\label{eq:compound}
    c_i(t) = h(t) \ast g_i(t)\,.
\end{equation}
We assume that the $c_i(t)$, ${i=0,\dots,L-1}$, are time-limited to the duration of the OFDM symbol's cyclic prefix, i.e., $c_i(t) = 0$ for $t \notin [0, T_{\textrm{cp}}]$. Therefore, there is no inter-symbol interference, allowing us to consider the signal model for a single OFDM symbol.

The received signal at the $i$th frequency band after conversion to baseband is given by
\begin{equation}
\label{eq:rec_sig}
	y_i(t) = x(t) \ast c_i(t) + q_i(t)\,,
\end{equation}
\noindent where $q_i(t)$ is low-pass filtered Gaussian noise. Here, we assumed that the mobile node and anchor are phased synchronized (cf. the remark at the end of this section for the signal model in the presence of the phase offset).
%
After conversion to the frequency-domain, the continuous-time model for the received signal $y_i(t)$ is given by
\begin{equation}
    \label{eq:freq_rx_sig}
       Y_i(\omega)  = 
       \begin{cases}
       X(\omega) C_i(\omega) + Q_i(\omega), & \omega \in \left[-\frac{B}{2}, \frac{B}{2}\right]
       \\
       \mbox{0}, & \text{otherwise}\,,
       \end{cases}
\end{equation}
\noindent where $C_i(\omega)$ is the compound Channel Frequency Response (CFR), and $X(\omega)$ and $Q_i(\omega)$ are the CTFTs of $x(t)$ and $q_i(t)$, respectively. 
\TA{Further, with slight abuse of notation ${C_i(\omega) = G_i(\omega)H_i(\omega)}$, where in $H_i(\omega)$ we implicitly take into account for the bandwith limitation effect of $G_i(\omega)$ on the CTFT of $h(t)$, and write $H_i(\omega)$ as}
\begin{equation}
\label{eq:chan_imp_bl}
    H_i(\omega) = \sum_{k=1}^{K} \alpha_k e^{-j(\omega_i+\omega)\tau_k}\,, 
    \quad  \omega \in \left[-\frac{B}{2}, \frac{B}{2}\right]\,.
\end{equation}

\squeezeup
\squeezeup
\subsection{Discrete Data Model}
\label{sc:discreteModel}
The receiver samples signal $y_i(t)$ with period $T_{\textrm s}=1/B$, performs packet detection, symbol synchronization, and removes the cyclic prefix. During the period of a single OFDM symbol, $N$ complex samples are collected, where $N$ is equal to the number of sub-carriers and $T_{\textrm{sym}} = NT_{\textrm s}$. Next, an $N$-point DFT is applied on the collected samples, and they are stacked in increasing order of DFT frequencies in $\by_i \in \bbC^{N}$. The discrete-time data model of the received signal (\ref{eq:freq_rx_sig}) can be written as
\begin{equation}
    \label{eq:data_model}
        \by_i = \diag(\bs) \bc_i + \bq_i\,,
\end{equation}
\noindent where ${\bq_i \sim \pazocal{CN}(\bzero, \sigma_{\mathrm{q}_i}^2 \bI_N)}$. The vector $\bc_i$ collects $N$ samples of the compound CFR at the subcarrier frequencies, and its entries are 
\begin{equation}
    \label{eq:compound_model1}
        [\bc_i]_n = \int_{0}^{T_{\textrm sym}} c_i(t)e^{-j\omega_{\textrm sc} n t}dt\,, \quad n = -\frac{N}{2}, \dots, \frac{N}{2}-1\,,
\end{equation}
\noindent where $\omega_{\textrm sc} = \frac{2\pi}{NT_{\textrm s}}$, and we assume that $N$ is an even number. Similarly, from (\ref{eq:compound}) we obtain (see \cite{730462} for details)
\begin{equation}
    \label{eq:compound_model2}
        \bc_i = \diag(\bg_i) \bh_i\,,
\end{equation}
\noindent where $\bg_i$ and $\bh_i$ collect samples of $G_i(\omega)$ and $H_i(\omega)$ at the subcarrier frequencies, respectively. We further refer to $\bh_i$ as the CSI vector and its entries are given as
\begin{equation}
    \label{eq:ch_dis1}
        H_i[n] = H_i\left(n\omega_{\textrm{sc}} \right), \quad n = -\frac{N}{2}, \dots, \frac{N}{2}-1\,.
\end{equation} 
\noindent We consider that the bands $\left\{\cW_i\right\}_{i=0}^{L-1}$ lie on a discrete frequency grid, i.e., $\omega_i = \omega_0 + n_i\omega_{\textrm sc}, i = 1, \dots, L-1$, where $n_i \in \bbN$. This is always the case in the WiFi standards \cite{lopez2019ieee}. Inserting the channel model (\ref{eq:chan_imp_bl}) into (\ref{eq:ch_dis1}) gives
\begin{equation}
    \label{eq:ch_dis2}
        H_i[n] = \sum_{k=1}^{K} \alpha_k e^{-jn_i\omega_{\textrm{sc}} \tau_k}e^{-jn\omega_{\textrm{sc}}\tau_k},
\end{equation}
\noindent where $e^{-j\omega_{0}\tau_k}$ is absorbed in $\alpha_k\,\forall\,k$. Then, $\bh_i$ can be written in a more compact form as
\begin{equation}
\label{eq:ch_model}
    \bh_i = \bM\bTheta_i\balpha\,,
\end{equation} 
where $\bM \in \bbC^{N\times K}$ is a Vandermonde matrix, given by
\begin{equation}
\label{eq:f_matrix}
\bM = 
\begin{bmatrix}
1 & 1 & \cdots & 1 \\
\upphi_1 & \upphi_2 & \cdots & \upphi_K\\
\vdots & \vdots & \ddots & \vdots \\ 
\upphi_1^{{N-1}} & \upphi_2^{N-1} & \cdots & \upphi_K^{N-1}
\end{bmatrix}
\,,\quad
\end{equation}
$\upphi_k = e^{-j\phi_k}$, and $\phi_k = \omega_{\textrm{sc}} \tau_k$ is the subcarrier-dependent phase shift introduced by the $k$th MPC. Likewise, $\bTheta_i$ is a diagonal matrix that collects the band dependent phase shifts introduced by the delays $\{\tau_k\}_{k=1}^K$, and ${\balpha = [\alpha_1,\dots,\alpha_K]^T \in \bbC^K}$. In view of the band positions on the frequency grid, we can write $\bTheta_i = \bPhi^{n_i}$, where ${\bPhi = \diag([\upphi_1 \cdots \upphi_K])}$.

\TA{We assume that none of the entries of $\bs$ or $\bg_i$ are zero or close to zero, so we can estimate the CSI from the data vector $\by_i$  using the classical LS estimation as $\bh_i = \diag^{-1}(\bs\odot\bg_i)\by_i$ \cite{edfors1998ofdm}. Then, from models (\ref{eq:data_model}) and (\ref{eq:compound_model2}), with a slight abuse of notation considering $\bq_i$, follows that $\bh_i$ satisfies the model}
\begin{equation}
    \label{eq:ch_est_model}
        \bh_i = \bM\bPhi^{n_i}\balpha + \bq_i\,.
\end{equation}
\noindent Here, we assume that the frequency response $\bg_i$ of the RF chains  is calibrated and known. Joint calibration and delay estimation is presented in \cite{9054034}. The training symbols $\bs$ typically have a constant magnitude by design, and we assume that the frequency responses of the receiver chains $\bg_i$ can be assumed almost flat for a single frequency band. Therefore $\bq_i$ is zero-mean white Gaussian distributed noise with covariance $\bR_{\mathrm{q}_i} = \sigma_{\mathrm{q}_i}^2\bI_N$. When the frequency responses of the RF chains are not flat, $\bq_i$ will be colored noise. However, its coloring is known and can be taken into account. We conclude this section with a remark on the influence of phase offset on the estimated channel model  (\ref{eq:ch_est_model}). 
	
\noindent \textbf{Remark.} If the mobile node and anchor are not phase synchronized, i.e., ${\psi_{\textrm  M, i} \not\approx \psi_{\textrm A, i}}$, the data model for the CSI collected at a mobile node becomes
\begin{equation}
    \label{eq:ch_model_ph}
        \bh_{\textrm M, i} = \bPsi_i\bh_i\,,
\end{equation}
\noindent where $\bPsi_i  = e^{-j \psi_i} \bI_N$ and $\psi_i = \psi_{\textrm M, i} - \psi_{\textrm A, i}$ is the unknown phase offset at the $i$th carrier frequency. The phase offset changes whenever the carrier frequency of the transceivers is changed. However, assuming that the transceiver is capable of Tx/Rx switching while keeping the phase lock loop (PLL) in-lock, $\psi_i$ stays the same for a fixed carrier frequency and has the opposite sign when estimated at the mobile node compared to the anchor. Using this property and assuming that the channel is reciprocal, we can write the model for the CSI collected at the anchor as $\bh_{\textrm A, i} = \bPsi_{i}^{*}\bh_i$. Now, the phase offset can be eliminated by taking the square-root of the point-wise product between collected CSIs as $\bh_{\textrm D,i} = (\bh_{\textrm M, i} \odot \bh_{\textrm A, i})^{1/2} = \pm \bh_i$, where the exponent is applied element-wise. Here, the square-root is used to avoid generation of additional unknown delays which are the result of inter-products between $\{\upphi_k\}_{k=1}^K$. The resulting measurements satisfy the model $\bh_{\textrm D, i} = \pm  \bM\bTheta_i\balpha$, where the ambiguity can be resolved by tracking the phase difference between multiple bands \cite{boer2020performance}.

\squeezeup
\section{Multiband delay estimation}
\label{sc:algorithm}
Given the CSI estimates $\bh_i$, $i=0,\dots,L-1$, the problem of ranging is to detect the LOS MPC and estimate its delay $\tau_1$. Then the range between the mobile node and an anchor is given by $d = \tau_1 c$, where $c$ denotes the speed of light. To do this accurately, all MPCs present in the channel need to be resolved and accordingly their delay and amplitude parameters $\{\tau_k, \alpha_k\}_{k=1}^K$ must be estimated. We start by stacking the CSI estimates $\bh_i$, $i=0,\dots,L-1$, into a multiband CSI vector $\bh = [\bh_0^T,\dots,\bh_{L-1}^T]^T \in \bbC^{NL \times 1}$. Using the model (\ref{eq:ch_est_model}), it follows that $\bh$ satisfies  
\begin{equation}
    \label{eq:multiband}
        \bh = \bA\balpha + \bq := 
 \begin{bmatrix} 
      \bM \\
      \bM\bPhi^{n_1}\\
      \vdots\\
      \bM\bPhi^{n_{L-1}}\\
  \end{bmatrix}\balpha + 
  \begin{bmatrix}
      \bq_{0} \\ \bq_{1} \\ \vdots \\ \bq_{L-1}
  \end{bmatrix} \,.
\end{equation}
If the band center frequencies $\omega_i$ are uniformly spaced, then matrix $\bA$ has a multiple shift-invariance structure and resembles the data model of Multiple Invariance ESPRIT \cite{swindlehurst1992multiple}, and this was exploited in our initial work \cite{kazaz2019time}. But also in the more general case, the overall structure present in (\ref{eq:multiband}) can be exploited to estimate the delay parameters $\{\tau_k\}_{k=1}^K$ from the phase shifts $\bphi = [\phi_1 \cdots \phi_K]$. These phase shifts are introduced over both subcarrier and frequency band apertures. The small aperture of the subcarriers promotes poor resolution but unambiguous estimation, while the very large aperture of the bands favors high resolution but ambiguous estimation of the delay parameters. We aim at an algorithm that will provide both high resolution and unambiguous delay estimates. To utilize all the structure present in the measurements, we formulate the multiband delay estimation as a multidimensional spectral estimation problem. We then propose an algorithm that estimates the delays $\{\tau_k\}_{k=1}^K$ by solving a weighted subspace fitting problem. After estimating the delays, the amplitudes $\{\alpha_k\}_{k=1}^K$ are estimated by solving a linear LS problem. 

\squeezeup
\squeezeup
\subsection{Algorithm Outline} 
\label{sc:algoritham_outline}
We first outline the key idea and the procedure for the estimation, and then introduce improvements to arrive at the final algorithm.

In subspace fitting methods, we would like to estimate the column span of $\bA$ in (\ref{eq:ch_est_model}). However, this ``signal subspace'' cannot be directly estimated from a single snapshot of the multiband CSI $\bh$. To restore the rank, we construct Hankel matrices $\bH_i$ of size $P\times Q$ from the vectors $\bh_i$, $i=0, \cdots, L-1$, as 
\begin{equation}
    \label{eq:hankel}
    \bH_i := \begin{bmatrix}
    H_i[0] & H_i[1] & \cdots & H_i[Q-1] \\
    H_i[1] & H_i[2] & \cdots &   H_i[Q] \\
    \vdots &  \vdots   &   \ddots     & \vdots  \\
    H_i[P-1] & H_i[P] & \cdots & H_i[N-1]
    \end{bmatrix} \,,
\end{equation}
\noindent where $P$ is a design parameter, and $Q = N-P+1$.
From (\ref{eq:ch_est_model}) and the shift-invariance structure present in $\bM$, the constructed Hankel matrices have the factorization 
\begin{equation} 
    \label{eq:hen_mat}
    \bH_i = \bM' \bPhi^{n_i} \bX + \bQ'_i \,,
\end{equation} 
where $\bM'$ is an $P\times K$ submatrix of $\bM$,
\[
\bX := [\balpha\, \quad \bPhi \balpha\, \quad \bPhi^2 \balpha\, \cdots \, \bPhi^{Q-1}
\balpha], \,
\]
and $\bQ'_i$ is a noise matrix with covariance $\bR_{\mathrm{q'}_i} = \sigma_{\mathrm{q'}_i}^2\bI_P$. Then we construct a block-row matrix $\bH$ of size $LP \times Q$ by stacking matrices $\bH_i$, ${i=0,\dots, L-1}$, as
\begin{equation}
    \label{eq:H} 
    \bH := \begin{bmatrix} \bH_0 \\ \bH_1 \\ \vdots\\ \bH_{L-1} \end{bmatrix}\,.
\end{equation}
\noindent The matrix $\bH$ preserves the shift-invariance properties of $\bh$ and has a factorization
\begin{equation} 
    \label{eq:H_fac} 
    \bH = \bA'(\bphi)\bX  + \bQ' := 
        \begin{bmatrix} 
          \bM' \\
          \bM' \bPhi^{n_1}\\
          \vdots\\
          \bM' \bPhi^{n_{L-1}}\\
       \end{bmatrix} \bX +
       \begin{bmatrix} 
          \bQ'_0\\
          \bQ'_1\\
          \vdots\\
          \bQ'_{L-1}\\
        \end{bmatrix}\,.
\end{equation}
\noindent Therefore, if we can choose the design parameter $P$ such that both $LP\ge K$ and $Q\ge K$ and if all factors in (\ref{eq:hen_mat}) are full rank, then $\bH$ has rank $K$, the number of MPCs present in the channel. This means that from the column span of $\bH$ we can estimate matrix $\bA'$ up to a $K \times K$ non-singular matrix $\bT$. In other words, we can write ${\bA' = \bU\bT^{-1}}$, where the columns of $\bU$ form a $K$-dimensional basis of the column space of $\bH$. 

The matrix $\bU$ can of course be estimated using a singular value decomposition (SVD) of $\bH$, and selecting the left singular vectors corresponding to the $K$ largest singular values $\{\lambda_j\}_{j=1}^K$. If the noise levels $\sigma_{\mathrm{q'}_i}^2$, $i=1,\dots,L-1$, are known and unequal, the blocks $\bH_i$ can be prewhitened prior to taking the SVD of $\bH$. \TA{The dimension $K$ can be estimated from the singular values using information-theoretic criteria \cite{wax1985detection}. In particular, in Section \ref{sc:real_exp} we find $K$ as the value $k \in \{0, 1, \dots, Q-1\}$ that minimizes modified minimum description length (MDL) criteria \cite{gaber2014study} given by 
\begin{equation} 
    \label{eq:mdl}
    \begin{aligned} \centering
         \mathrm{MDL}(k) =& -(D-k)D \cdot \log \frac{\prod_{j=k+1}^{D}\lambda_j^{1/(D-k)}}{\frac{1}{D-k}\sum_{j=k+1}^D\lambda_j} \\
         &+ k(2D-k)\cdot\log(D)/4+k\,,
    \end{aligned}
\end{equation}
\noindent where $D = Q-1$.}

The estimation of $\bphi$ from $\bH$ is based on exploiting the shift invariance structure present in $\bA'$ and $\bU$. Accounting for the errors introduced during estimation of  $\bU$, we can write $\bA'(\bphi) \approx \hat{\bU}\bT^{-1}$. Now, to estimate $\bphi$, we formulate the subspace fitting problem
\begin{equation} 
    \label{eq:sub_fit_prob1}
    \centering
    \xhat{\bphi}, \hat{\bT} = \argminA_{\bphi, \bT} \left\|
    \hat{\bU} - \bA'(\bphi) \bT \right\|_{F}^2\,.
\end{equation}
\noindent The problem of minimizing the cost function in (\ref{eq:sub_fit_prob1}) is a nonlinear LS problem (NLS). It is easy to see that for the optimal $\bphi$, the optimal $\bT$ must satisfy ${\bT = \bA'^{\dagger}(\bphi)\hat{\bU}}$. Therefore, this problem can be further recast into a separable nonlinear LS (SNLS) problem \cite{golub2003separable},
\begin{equation} 
    \label{eq:sub_fit_est1}
    \begin{aligned} \centering 
        \xhat{\bphi} &= \argminA_{\bphi} J(\bphi)\,, \\
           J(\bphi) &= \tr\left(\bP_{\bA'}^{\perp}(\bphi)\hat{\bU}\hat{\bU}^{H}\right)\,,
    \end{aligned}
\end{equation}
\noindent where $\bP_{\bA'}^{\perp}(\bphi) = \bI - \bP_{\bA'}$ and $\bP_{\bA'} = \bA'\bA'^{\dagger}$ is a projection onto the column span of $\bA'$. This reformulation reduces the dimension of the parameter space and also results in a better-conditioned problem, which can be efficiently solved using iterative optimization methods such as variable projection or the Levenberg-Marquardt (LM) \cite{golub2003separable}. We use the LM method, where good initialization of the algorithm is obtained by the multiresolution (MR) delay estimation algorithm \cite{kazaz2019multiresolution}. With this initialization, the LM method converges very fast, typically within five steps for moderate signal-to-noise ratios (SNRs) as shown in Section \ref{sc:resconv}.

\squeezeup
\subsection{Weighting} 
\label{sc:weightening}
The dominant sources of estimation errors in (\ref{eq:sub_fit_prob1}) are caused by perturbations of the subspace estimates. The estimated singular vectors in $\hat{\bU}$ are each perturbed differently. Thus, the estimator based on unweighted subspace fitting is not statistically efficient, and it is sensitive to noise. These errors can be reduced by introducing an appropriate column weighting in the cost function $(\ref{eq:sub_fit_est1})$, \cite{swindlehurst1993performance}. Therefore, to improve estimation and to penalize subspace perturbations errors, we estimate $\bphi$ by solving the following weighted subspace fitting problem
\begin{equation} 
    \label{eq:sub_fit_prob2}
    \centering
    \xhat{\bphi}, \hat{\bT} = \argminA_{\bphi, \bT} \left\| \left(
    \hat{\bU} - \bA'(\bphi) \bT \right)\TA{\bW} \right\|_{F}^2\,,
\end{equation}
\noindent where \TA{$\bW$} is a $K \times K$ matrix. Similar as in (\ref{eq:sub_fit_est1}), this problem can be recast to the SNLS problem with a cost function $J(\bphi) = \tr (\bP_{\bA'}^{\perp}(\bphi)\hat{\bU}\TA{\bW}^2\hat{\bU}^{H})$, and the same initialization and optimization methods can be applied to find the solution. The matrix \TA{$\bW$} is assumed to be positive definite and Hermitian, and its role is to whiten perturbations of the singular vectors in $\hat{\bU}$. A good choice for \TA{$\bW$} is given in \cite{swindlehurst1993performance} as
\begin{equation} 
    \label{eq:row_col_weig}
    \centering
    \TA{\bW} = \hat{\bLam}_{\textrm s} - \hat{\sigma}^2 \bI_K \,,
\end{equation}
\noindent where $\hat{\bLam}_{\textrm s}$ is a diagonal matrix that collects the $K$ largest squared singular values of $\bH$ and $\hat{\sigma}^2$ is the estimated noise power. The noise power $\hat{\sigma}^2$
follows from the noise levels $\sigma_{\mathrm{q'}_i}^2$, $i=1,\dots, L-1$. If these are unequal, we would prewhiten the blocks $\bH_i$ prior to taking the SVD of $\bH$.

\squeezeup
\subsection{Data Extensions}
In this section, we discuss techniques for extending the data matrix $\bH$ if multiple channel measurements are available or if subcarrier frequencies of a multiband training signal satisfy a centro-symmetric configuration.

\subsubsection{Multiple Snapshots}
So far, we have assumed that the CSI is collected only once during the channel coherence time. However, the coherence time of common multipath radio channels is much longer than the duration of training signals. For example, the indoor radio channel that characterizes propagation of WiFi signals in the 2.4 GHz frequency band between anchors and pedestrians with a velocity of 1 m/s has a coherence time of approximately 53 ms. Now, assuming that a WiFi training signal with a duration of 40 $\mathrm{\mu}$s is used to estimate the CSI, then at least 50 snapshots of CSI can be collected during the coherence time.

Let us assume that $M$ snapshots of multiband CSI (\ref{eq:multiband}) are collected during the coherence time. These measurements satisfy the model 
\begin{equation}
\label{eq:multiband_ms}
    \bh^{(m)} = \bA\balpha^{(m)} + \bq^{(m)} \,, \quad m = 1, \dots, M\,.
\end{equation}
where $\balpha^{(m)}$ collects the complex amplitudes of the MPCs. Similar as in the single snapshot case, from every snapshot $\bh^{(m)}$ a block Hankel matrix $\bH^{(m)}$ is formed as shown in Section \ref{sc:algoritham_outline}. \TA{We assume that the delays $\{\tau_k^{(m)}\}_{k=1}^K$ of MPCs stay the same during the coherence time. On the other hand, we assume that the amplitudes $\alpha^{(m)}$ are independent complex Gaussian random variables that vary with time where their mean magnitudes stays constant.} Similar as in (\ref{eq:H_fac}), the matrix $\bH^{(m)}$ satisfies the model ${\bH^{(m)} := \bA'\bX^{(m)} + \bQ'^{(m)}}$, where now ${\bX^{(m)} := [\balpha^{(m)}\, \quad \bPhi \balpha^{(m)}\, \cdots \, \bPhi^{Q-1}\balpha^{(m)}]}$,  and $\bQ'^{(m)}$ represents the noise matrix of the $m$th snapshot. The matrices $\bH^{(m)}$, $m = 1, \dots, M$, have the same column subspace and from them an extended $LP \times QM$ data matrix is constructed as
\begin{equation} 
    \label{eq:ext_data}
    \centering
    \bcH := [\bH^{(1)}\, \quad \bH^{(2)} \, \cdots \, \bH^{(M)}]\,.
\end{equation}
\noindent The matrix $\bcH$ has a factorisation 
\begin{equation} 
    \label{eq:ext_data_fac}
    \centering
    \bcH = \bA' \bcX + \bcQ\,, 
\end{equation}
\noindent where ${\bcX := [\bX^{(1)}\, \cdots \, \bX^{(M)}]}$ and ${\bcQ := [\bQ'^{(1)}\, \cdots \, \bQ'^{(M)}]}$. The estimation of $\bphi$ from $\bcH$ proceeds as described in Section \ref{sc:weightening}. However, the number of columns in the data matrix is now increased, which provides improvement of estimation accuracy in terms of noise. Multiple snapshots also enables the opportunity to increase the number of rows in $\bH^{(m)}$ as now the number of columns $Q$, necessary to restore the dimension of the signal subspace, can be smaller: $Q \ge \text{max}(1, K+1-M)$. Increasing the number of rows in $\bH^{(m)}$ increases the frequency aperture and leads to improved delay resolution.

\subsubsection{Forward-Backward Averaging}
\label{sc:fb}
Another technique to extend the data matrix is known as forward-backward (FB) averaging \cite{pillai1989forward}. This technique can only be applied when multiband CSI is collected on a \textit{centro-symmetric} set of frequencies. Let the central frequency of the set of probed frequencies $\cW_i$, ${i=0, \dots, L-1}$, be defined as ${\omega_{\textrm c} = (\omega_{L-1}+\omega_{0})/2}$. A set of frequencies is centro-symmetric if for any frequency in the set there is a corresponding frequency located in the opposite direction and equidistant with respect to the central frequency of the set. If these constraints are satisfied, then FB averaging can be applied by exploiting the structure of $\bA$ in (\ref{eq:multiband}) and the fact that the ${\{\upphi_k\}}_{k=1}^K$ are on the unit circle. Let $\bPi$ denote the $LP \times LP$ exchange matrix that reverses the ordering of the rows, then it is seen that $\bPi\bA^* = \bA \bUpsilon$, for some unitary diagonal matrix $\bUpsilon$ related to $\bPhi$. In particular, $\bA$ and $\bPi\bA^*$ have the same column span. 

Thus, we can construct the forward-backward averaged multiple snapshot data matrix as
\begin{equation} 
    \label{eq:ext_fb}
    \centering
    \bcH_{\textrm e} := [\bcH \quad  \bPi \bcH^{\ast}]\,, 
\end{equation}
\noindent of size $LP \times 2QM$. Then, $\bcH_e$ has a factorisation
\begin{equation} 
    \label{eq:ext_fb_fac}
    \centering
    \bcH_{\textrm e} = \bA' \bcX_{\textrm e} + \bcQ_{\textrm e} := \bA' [\bcX \quad \bUpsilon \bcX^{\ast}] + [\bcQ \quad  \bPi\bcQ^{\ast}]\,.  
\end{equation}
\noindent Thus, the FB averaging doubles the number of columns of the data matrix, which leads to improved accuracy. It also provides the opportunity to increase the number of rows in $\bH^{(m)}$, as now the number of columns $Q$ necessary to restore the dimension of the signal subspace is half of what it used to be.
The estimation of $\btau$ from the extended data matrix proceeds as described in Section \ref{sc:weightening}.

\squeezeup
\subsection{Noise Reduction}
\label{sc:nr}
The Hankel matrices $\bH_i$, ${i=0,\dots, L-1}$, stacked in $\bH$, all have the same $K$-dimensional basis for their column spaces, i.e., the column span of $\bM'$. 
Instead of stacking the $\bH_i$ vertically into $\bH$, we can stack them horizontally. This allows us to obtain a good estimate of that basis. 

We consider the general case, and first, we exploit the structure of $\bM$ in (\ref{eq:f_matrix}), to apply FB averaging on each of $\bH_i$. The FB averaged multiple snapshot data matrix for the $i$th band is defined as ${\bcH_{{\textrm e}, i} := [\bcH_i\, \quad \, \bPi'\bcH_i^{\ast}]}$, where 
\begin{equation} 
    \label{eq:row_com_M}
    \centering
    \bcH_i := [\bH_i^{(1)}\, \dots \, \bH_i^{(M)}]\,,
\end{equation}
\noindent $\bH_i^{(m)}$ is Hankel matrix formed from CSI collected in the $i$th band at the $m$th snapshot, and $\bPi$ is the ${P \times P}$ exchange matrix. To estimate the basis, we construct 
    \begin{equation} 
        \label{eq:row_com}
        \centering
        \bcH_{\textrm r} := [\bcH_{\textrm e, 0}\, \quad \, \bcH_{\textrm e, 1} \, \cdots \, \bcH_{\textrm e, L-1}]\,,
    \end{equation}
\noindent which has a factorisation 
    \begin{equation} 
        \label{eq:row_comb_fac}
        \begin{aligned} \centering
            \bcH_{\textrm r} &= \bM\bcX_{\textrm r} + \bcQ_{\textrm r}\,\\
            &:= \bM [\bcX_{\textrm e, 0}\, \cdots \, \bTheta_{L-1} \bcX_{\textrm e, L-1}] + [\bcQ_{\textrm e, 0}\, \cdots \, \bcQ_{\textrm e, L-1}]\,.
       \end{aligned}
    \end{equation}
After computing the SVD of $\bcH_{\textrm r}$, let matrix $\hat{\bU}_r$ contain the $K$ dominant left singular vectors, i.e., the estimated basis for the column span of $\bM'$. 

Moving back to the vertically stacked data matrix $\bcH_{\textrm e}$, the noise in this matrix can be reduced by projecting each of its blocks onto the low dimensional column span of $\hat{\bU}_r$:
    \begin{equation}
        \label{eq:projected_H} 
        \nonumber 
        \bcH_{\textrm p} = \left(\bI_L \otimes \bP_{\bU_r}\right) \bcH_{\textrm e}\,,
    \end{equation} 
where $\bP_{\bU_{\textrm r}} = \hat{\bU}_{\textrm r}\hat{\bU}_{\textrm r}^H$. The projected data matrix $\bcH_{\textrm p}$ has a factorisation
    \begin{equation} 
    \label{eq:cb_H_fac} 
    \bcH_{\textrm p} = \bA' \bcX_{\textrm e}  + \bcQ_{\textrm p} := 
        \begin{bmatrix} 
          \bM' \\
          \bM' \bPhi^{n_1}\\
          \vdots\\
          \bM' \bPhi^{n_{L-1}}\\
       \end{bmatrix} \bcX_{\textrm e} +
       \begin{bmatrix} 
          \bP_{\bU_{\textrm r}} \bcQ_{\textrm e, 0} \\
          \bP_{\bU_{\textrm r}} \bcQ_{\textrm e, 1} \\
          \vdots\\
          \bP_{\bU_{\textrm r}} \bcQ_{\textrm e, L-1}\\
        \end{bmatrix}\,.
    \end{equation}
The column space of the matrix $\bcH_{\textrm p}$ has the same structure as the column space of $\bcH_{\textrm e}$. However, the noise matrices $\bcQ_{\textrm e, i}$, $i=0,\dots,L-1$, are projected onto the lower dimensional subspace, which improves accuracy. The estimation of $\btau$ from $\bcH_{\textrm p}$ proceeds as described in Section  \ref{sc:weightening}.

\squeezeup
\subsection{Estimation of Amplitudes and Algorithm Summary}
After estimation of the delays $\btau$, the amplitudes $\balpha^{(m)}$ (if they are of interest) can be found as the LS solution to (\ref{eq:multiband_ms}), that is
\begin{equation}
\label{eq:amp_est}
    \hat{\balpha}^{(m)} = \hat{\bA}^{\dagger}\bh^{(m)} \,, \quad m = 1, \dots, M\,,
\end{equation}
\noindent where $\hat{\bA}$ is constructed based on model (\ref{eq:multiband}) using $\hat{\btau}$. 

A summary of the resulting Multiband Weighted Delay Estimation (MBWDE) algorithm is shown as Algorithm \ref{alg.mdwde}. With the input $\hat{\btau}_{\rm MR}$ we denote an initial estimate of $\btau$ obtained using the related multiresolution delay estimation algorithm \cite{kazaz2019multiresolution}. The abstract routine $\FCons(\cdot)$ points to the construction of $\bA$ or $\bA'$ from $\btau$ (via $\bPhi$) in (\ref{eq:multiband}) or (\ref{eq:H_fac}), respectively. $\TSVD$ refers to the truncated SVD (truncating at rank $K$). The remaining parts of the summary are self-explanatory.   

\begin{algorithm}[!t]
    \setstretch{1.1}
    \SetAlgoLined
    \KwInput{$\hat{\btau}_{\rm MR},N, P, K, \{\bh_i^{(m)}:m=1,\dots, M \}_{i=0}^{L-1}$}
    \KwOutput{$ \hat{\btau}, \hat{\balpha}^{(m)}$}
    $Q \larrow N-P+1$\;
    $\bH_i^{(m)} \larrow \FHank(\bh_i^{(m)}, P, Q), \forall i, m$; \hfill (\ref{eq:hankel})\\
    $\bcH \larrow [\bH^{(1)}\, \dots \, \bH^{(M)}], \forall i$; \hfill (\ref{eq:ext_data})\\
    $\bcH_{\rm e} \larrow  \bcH$; \hfill \\

    \If {ForwardBackward}{
        $\bcH_{\rm e} \larrow [\bcH \quad  \bPi \bcH^{\ast}]$; \hfill \quad (\ref{eq:ext_fb})\\
    }
    
    \If {NoiseReduction}{
        $\bcH_i \larrow [\bH_i^{(1)}\, \dots \, \bH_i^{(M)}], \forall i$; \hfill (\ref{eq:row_com_M})\\
        $\bcH_{\mathrm{e},i}  \larrow [\bcH_i \quad  \bPi' \bcH_{i}^{\ast}]$;\\
        $\bcH_{\rm r} \larrow [\bcH_{\textrm e, 0}\, \quad \, \bcH_{\textrm e, 1} \, \cdots \, \bcH_{\textrm e, L-1}]$; \hfill (\ref{eq:row_com})\\
        $\hat{\bU}_{\rm r}  \larrow \TSVD(\bcH_{\rm r},K)$\;
        $\bP_{\bU_{\rm r}} \larrow \hat{\bU}_{\rm r}\hat{\bU}_{\rm r}^H$\;
        $\bcH_{\rm e} \larrow \left(\bI_L \otimes \bP_{\bU_{\rm r}}\right) \bcH_{\rm e}$\;
    }
    $\{\hat{\bU}, \hat{\bLam}_{\textrm s}, \hat{\sigma}^2\} \larrow \TSVD(\bcH_{\rm e},K)$\;
    $\TA{\bW} \larrow \hat{\bLam}_{\textrm s} - \hat{\sigma}^2 \bI_K$; \hfill
    (\ref{eq:row_col_weig})\\
    $\hat{\bA}'_{\rm MR} \larrow \FCons \left(\hat{\btau}_{\rm MR}\right)$; \hfill (\ref{eq:multiband})\\
    $\hat{\btau} \larrow \FWeigSubFit(\hat{\bA}'_{\rm MR}, \hat{\bU}, \TA{\bW})$; \hfill (\ref{eq:sub_fit_prob2})
    \\[1.5ex]
    $\bh^{(m)} \larrow [\bh_1^{(m)T} \dots \bh_{L-1}^{(m)T}]^T,\forall m$\;
    $\hat{\bA} \larrow \FCons \left(\hat{\btau}\right)$; \hfill (\ref{eq:multiband})\\
    $\hat{\balpha}^{(m)} \larrow \FLS(\hat{\bA}, \bh^{(m)}), \forall m$; \hfill (\ref{eq:amp_est})
    \caption{Multiband Weighted Delay Estimation}
    \label{alg.mdwde}
\end{algorithm}


\section{Cram\'er-Rao Bound}
\label{sc:crb}
In this section, we derive the CRB for the model (\ref{eq:multiband_ms}), which sets a lower bound on the error covariance matrix of any unbiased estimator \cite{kay1998estimation}. After that, we analyze the effects of wireless system parameters, e.g., bandwidth, number of CSI measurements, and band selection, on the CRB.

The mean square error (MSE) of the estimated delays, when only errors due to the variance of the estimator are present \cite{kay1998estimation}, is defined as 
\begin{equation}
    \label{eq:mse}
    \textrm{MSE}(\hat{\btau}) := \bbE\{(\hat{\btau} - \btau)^2\} = \text{var}(\hat{\btau})\,,
\end{equation}
where $\text{var}(\hat{\btau})$ is the variance of the estimates. 

Let us assume that all MPCs are resolved and that the bias can be ignored, then the covariance matrix of the delay estimation errors and its lower bound are defined as
\begin{linenomath*}
\begin{align}
    \label{eq:crlb} 
    \bC_{\hat{\btau}} := \bbE\{(&\hat{\btau} - \btau)(\hat{\btau} - \btau)^T\} \succcurlyeq \text{CRB}(\btau)\,,\\
    &\text{CRB}(\btau) := \bF^{-1}\,,
\end{align} 
\end{linenomath*}
\noindent where $\hat{\btau}$ are estimated delays, $\bbE(\cdot)$ is the statistical expectation operator with respect to the squared error, $\bF$ is the Fisher's Information Matrix (FIM), and relationship $\bC_{\hat{\btau}} \succcurlyeq \text{CRB}(\btau)$ implies that the matrix $\bC_{\hat{\btau}}-\text{CRB}(\btau)$ is positive semidefinite. The entries on the diagonal of $\bC_{\hat{\btau}}$ are equal to the variances of the estimated delays $\text{var}(\hat{\btau})$.


The data model (\ref{eq:multiband_ms}) is familiar from array signal processing, and the FIM and the CRB for DOA estimation are derived in  \cite{stoica1989music}. We can readily adapt these results to the problem of delay estimation by making the following assumptions:
 \begin{enumerate}
      \item[(\textbf{A1})] The noise $\bq^{(m)}$ in the model (\ref{eq:multiband_ms}) is zero-mean circularly-symmetric Gaussian with covariance $\bR_{\textrm{q}} = \sigma_{\textrm{q}}^2 \bI_{LN}$. This assumption is satisfied when the transceivers have equal gain in all bands and the training symbols $\bs$ have a constant magnitude.
      
      \item[(\textbf{A2})] The amplitudes of the MPCs are assumed to be circularly symmetric complex Gaussian random variables, i.e., ${\alpha_k \sim \pazocal{CN}(0, \sigma_{\alpha,k}^2)}$, ${k=1,\dots, K}$, with covariance matrix $\bR_{\alpha}$. Thus, the magnitudes of the MPCs are Rayleigh distributed, and we assume that they have an exponentially decaying power-delay profile.
      
      \item[(\textbf{A3})] \TA{The FIM matrix given in (\ref{eq:fim1}) is non-singular and the CRB can be computed by taking its inverse. The validity of this assumption depends on the delay separation between MPCs with respect to the system bandwidth \cite{rottenberg2020performance}. As a rule of thumb, we say that matrix $\bF$ will become rank deficient if the delay separation of two MPCs is much smaller than the inverse of the total system bandwidth, i.e., much smaller than $1/B$. In numerical experiments presented in Section \ref{sc:resconv} we see that for $B =$ 80 MHz, this assumption is satisfied even when delay separation of MPCs is 125 times smaller than $1/B$.}

      \item[(\textbf{A4})] The MPCs and noise are temporally uncorrelated.
 \end{enumerate}
 
Based on the above assumptions, the FIM for the delay parameters, conditioned on the path amplitudes, is given as
\begin{equation} 
\label{eq:fim1} 
   \bF =
   \dfrac{2M}{\sigma_q^2} \text{Re} \left\{ \bD^H
   \bP_{\bA}^\perp\bD \odot
   \bR_{\alpha} \right\}\,,
\end{equation} 
\noindent where
\begin{equation} 
\label{eq:derv_steering} 
   \bD = \left[ \dfrac{\partial \ba(\tau_1)}{\partial \tau_1}, \dots, \dfrac{\partial
   \ba(\tau_K)}{\partial \tau_K} \right]\,,
\end{equation} 
\noindent $\ba(\tau_k)$ is the $k$th column of $\bA$, ${\bP_{\bA}^\perp = \bI_{LN} - \bP_{\bA}}$, and ${\bP_{\bA} = \bA (\bA^H \bA)^{-1}\bA^H}$. To gain further insights in the CRB we partition the FIM in terms associated to the delays of MPCs and their coupling with other multipath parameters, and write it in the following explicit form as
\begin{equation} 
\label{eq:fim2} 
   \bF =
   \dfrac{2M}{\sigma_q^2} \text{Re} \{ \underbrace{\bD^H\bD \odot
   \bR_{\alpha}}_{\substack{\text{Partition of FIM} \\ \text{of delay parameters}}} -   \underbrace{\bD^H \bP_{\bA} \bD \odot
   \bR_{\alpha}}_{\substack{\text{Partition of FIM} \\ \text{of coupled parameters}}}\}\,.
\end{equation} 
    \noindent We can make the following observations.
 \begin{itemize}
    \item The CRB depends on the delays $\btau$, frequency band selection $\{\cW_i\}_{i=0}^{L-1}$ through $\bA$ and $\bD$, and correlation between amplitudes $\balpha$ through $\bR_{\alpha}$.
    \item The first term in the FIM represents the effects of the delays $\btau$ on the estimation error, and is equivalent to the FIM for delay estimation in the additive white Gaussian noise channel when there is no multipath propagation. 
    \item The second term represents the effects of coupling between parameters $\btau$ and $\balpha$ on the estimation error of delays $\btau$. This term is always non-negative, and it will increase the CRB except when the parameters are decoupled. An increase of the CRB due to coupling of the parameters depends on the conditioning of matrix $\bA^H \bA$, and it will be low when this matrix is well-conditioned.
 \end{itemize}
 
Unfortunately, these observations do not intuitively interpret the impact of band selection $\{\cW_i\}_{i=0}^{L-1}$ on the CRB. To arrive at a more interpretable expression for the CRB, we will make the additional assumption that matrix $\bR_{\alpha}=\diag[\sigma_{\alpha,k}^2, k=1\dots K]$ is diagonal. This assumption holds for \textit{wide sense stationary uncorrelated scattering} (WSSUS) channels due to uncorrelated scattering. Then using (\ref{eq:fim1}), we can write the closed-form expression for the CRB delay estimates of the $k$th MPC as
\begin{equation} 
\label{eq:var_los_crb} 
   \text{CRB}(\hat{\tau}_k) = \dfrac{1}{2M \cdot \text{SNR}_k} b^{-1}(\tau_k)\,,
\end{equation} 
\noindent where $\text{SNR}_k = \sigma_{\alpha,k}^2/\sigma_{\mathrm{q}}^2$, ${b(\tau_k) = \bd^H(\tau_k)(\bI_{LN}-\bP_{\bA})\bd(\tau_k)}$ and $\bd(\tau_k)$ is the $k$th column of $\bD$. This expression shows that the CRB is inversely proportional to the number of snapshots $M$, $\text{SNR}_k$ and the scalar $b(\tau_k)$, where $b(\tau_k)$ depends on the coupling between the parameters. If we ignore the effects of the coupling, then ${\bd^H(\tau_k)\bP_{\bA}\bd(\tau_k) = 0}$, and (\ref{eq:var_los_crb}) reduces to the CRB for delay estimation in AWGN channels \cite{driusso2014performance}. The scalar  $b(\tau_k)$ then can be written as ${b(\tau_k) = \sum_{n \in \cS} (\omega_{sc}n)^2}$, where $\cS$ is the index set of all used subcarriers of all frequency bands. It is defined as $\cS = \bigcup_{i=0}^{L-1} \cS_i$, where $\cS_i = \{n \in \bbZ \mid n_{c,i} - \frac{N}{2} \leq n < n_{c,i} + \frac{N}{2} \}$, $n_{c, i} = n_i - \frac{n_{L-1}-n_0}{2}$, and $n_i = \frac{\omega_i-\omega_0}{\omega_{sc}}$, $i= 0, \dots, L-1$. 
Now, it is easy to see that the CRB (\ref{eq:var_los_crb}) can be reduced by collecting the CSI over a large frequency aperture. However, the results of real data experiments show that a large frequency aperture introduces modeling errors caused by frequency dependency of multipath channels \cite{malik2007frequency}. \TA{Therefore the bands need to be selected carefully, and this is further discussed in Section \ref{sc:real_bandselection}.}

\section{Numerical Experiments}
\label{sc:num_exp}
This section presents numerical results that illustrate the performance of the MBWDE algorithm. We first describe the simulation setup and then compare different variants of the algorithm and study how the trade-offs among design and system parameters impact the performance. Lastly, we compare the performance of the algorithm against several other algorithms. The results show that the algorithm is asymptotically efficient, achieves the CRB, and improves the resolution of delay estimation with respect to the bandwidth of training signals.

In the simulations, we consider delay estimation using IEEE 802.11be transceivers. Although the IEEE 802.11be standard is in a preliminary phase, its main candidate features are already known \cite{lopez2019ieee}. In particular of interest to us is that it will enable multiband operation at 2.4, 5, and 6 GHz as shown in Fig. \ref{fig:fig_ilu3}. At 6 GHz, the RF spectrum from 5.925 to 7.125 GHz will be allocated for primary 20, 40, 80, and 160 MHz channels and their contiguous and non-contiguous combinations. The large bandwidth allocated at the 6 GHz band offers a great opportunity for localization.

\begin{figure}[t!] 
    \centering
    \includegraphics[width=0.85\columnwidth]{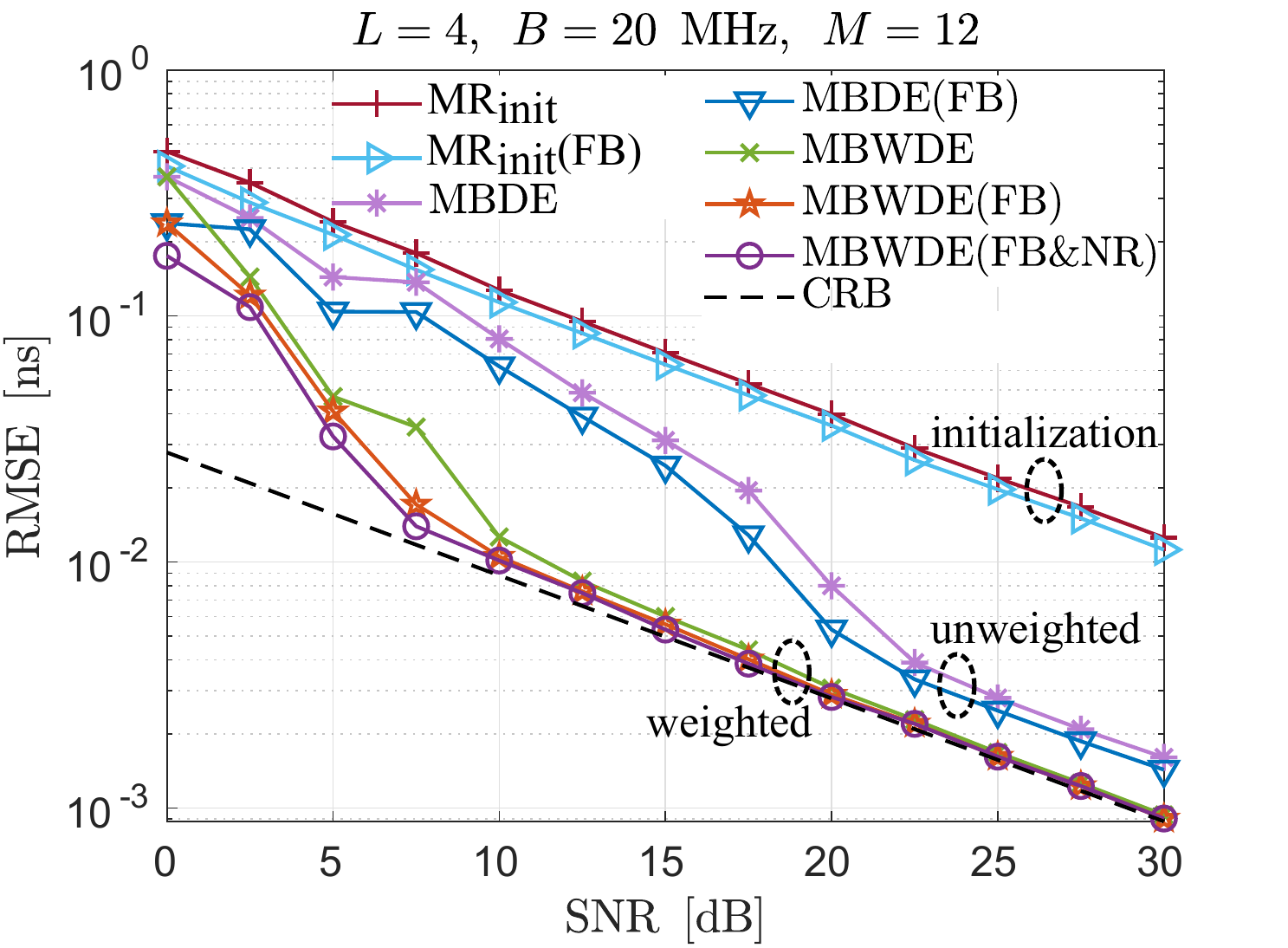}
    \caption{RMSE of delay estimation for different variants of MBDE and MBWDE algorithms.}
\label{fig:fig_res1}
\end{figure}
\begin{figure}[t!] 
    \centering
    \includegraphics[width=0.85\columnwidth]{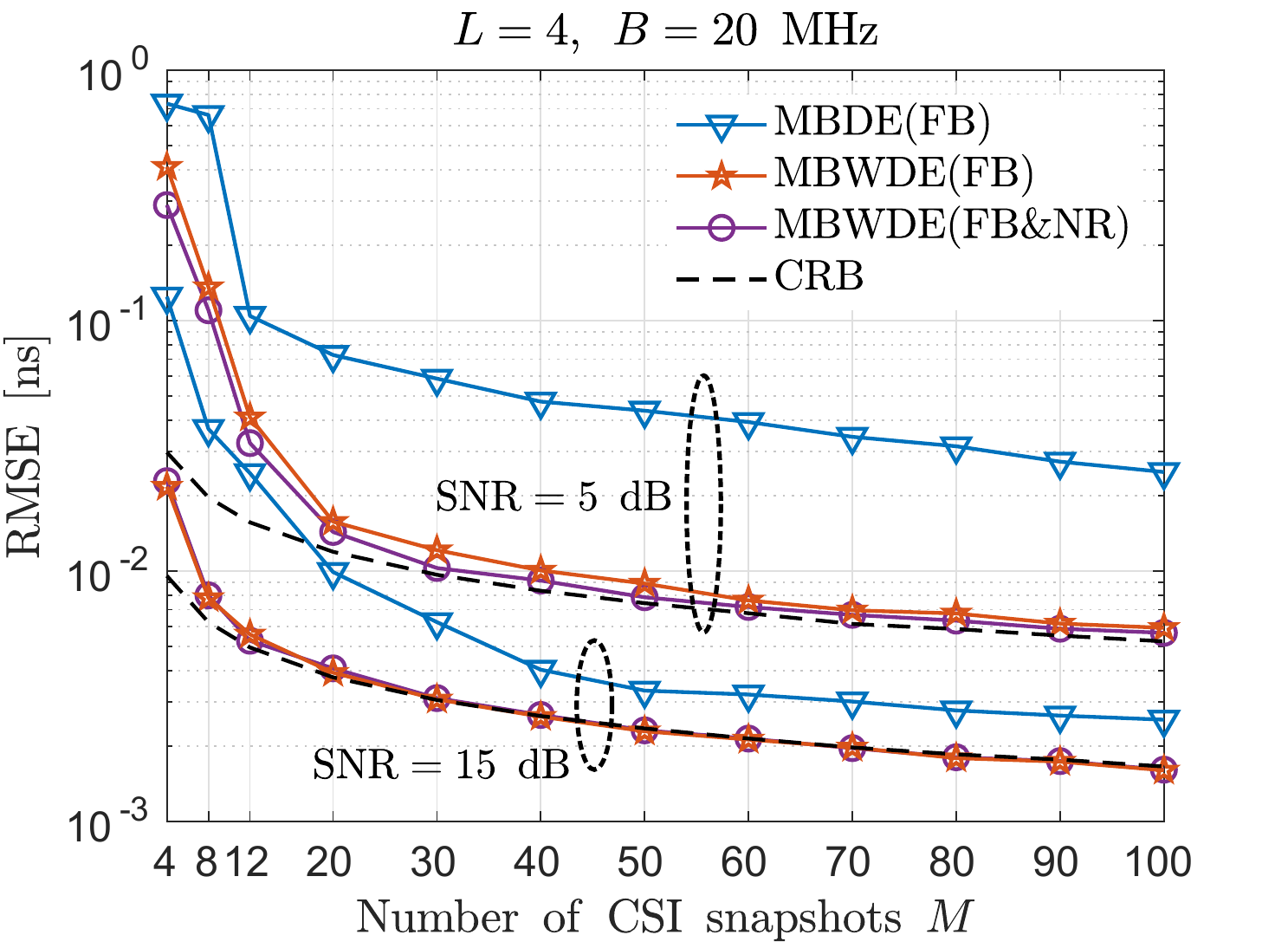}
    \caption{Influence of the number of CSI snapshots $M$ on the performance of delay estimation.}\label{fig:fig_res5}
\end{figure}

\begin{figure*}[t!] 
    \centering
    \subfloat[]{\includegraphics[width=0.85\columnwidth]{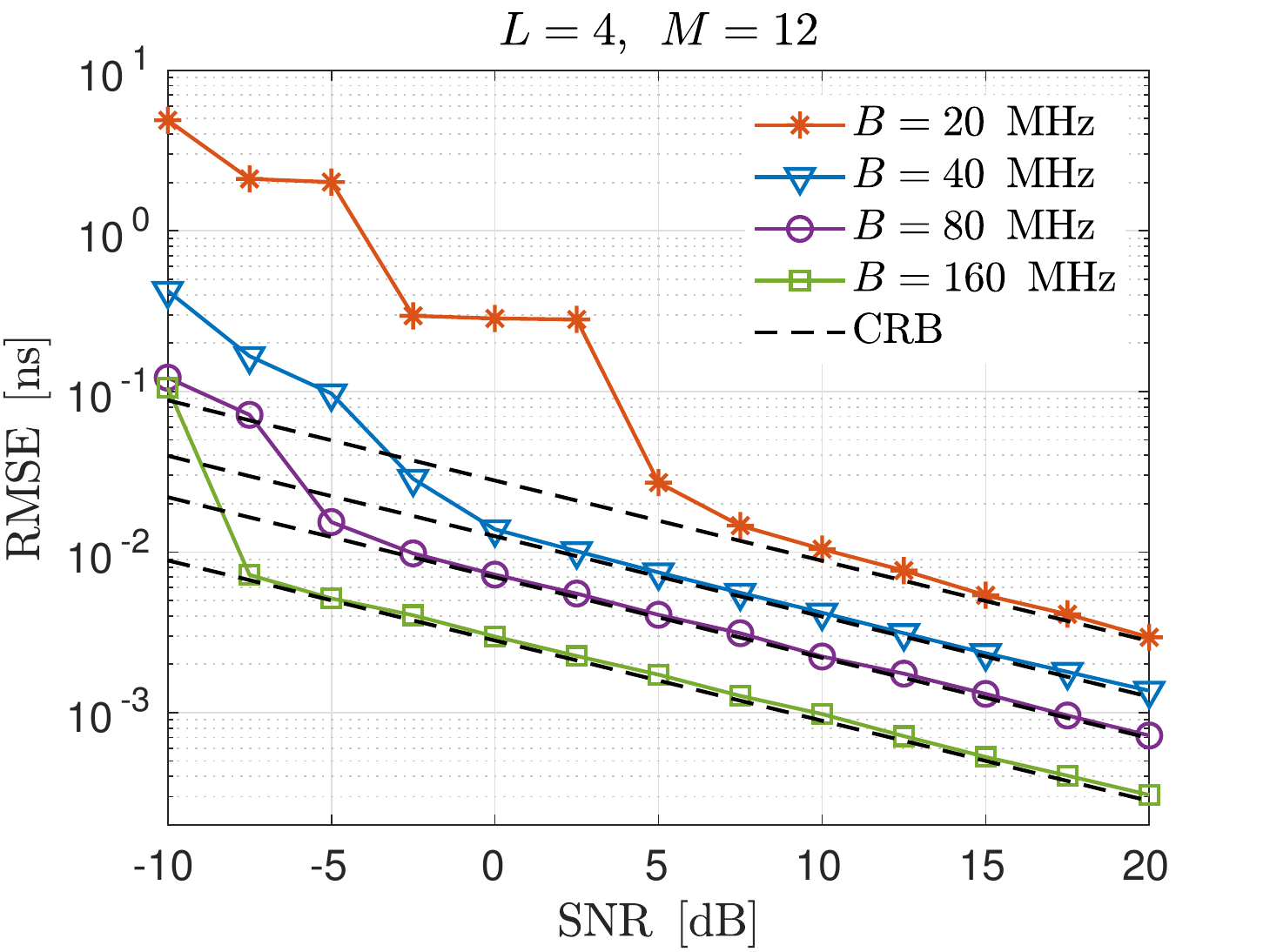}\label{fig:figure_res6a}}
    \hfil
    \subfloat[]{\includegraphics[width=0.85\columnwidth]{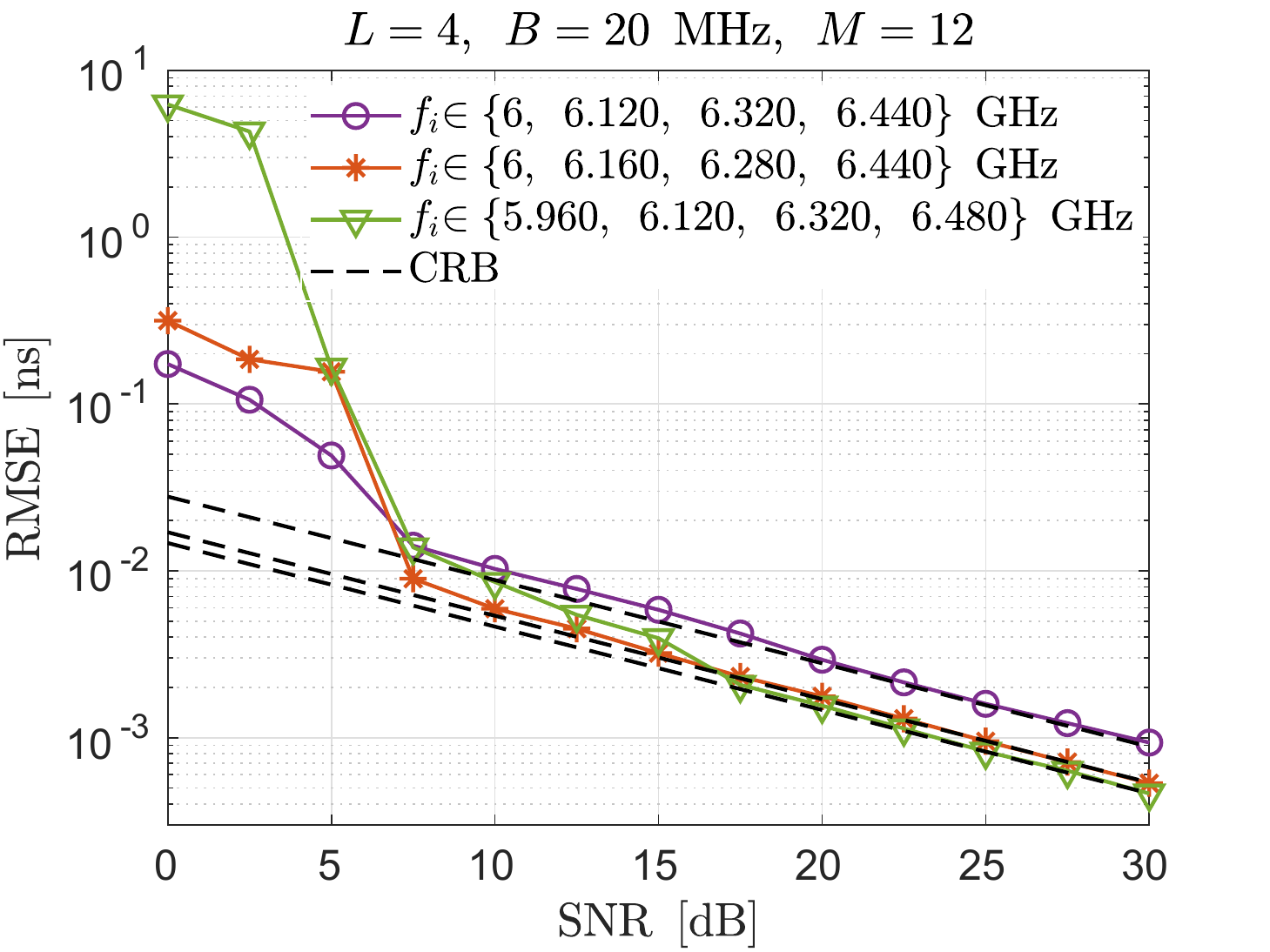}\label{fig:figure_res6b}}
    \caption{(a) Impact of the bandwidth $B$ of training signals on the RMSE. (b) Impact of the choice of carrier frequencies of the bands on the RMSE.}
\label{fig:figure6}
\end{figure*}

In the default setup, we consider that CSIs are collected using OFDM training signals with subcarrier spacing $\omega_{sc} = \mathrm{78.125}$ kHz and bandwidth of $B=\mathrm{20}$ MHz at $L=\mathrm{4}$ bands, with central frequencies $\mathrm{\{6, 6.120, 6.320, 6.440\}}$ GHz. This corresponds to probing the channel using $20$ MHz wide extremely high throughput long training fields (EHT-LTF) described in the standard. We consider that $M=12$ CSI snapshots are collected within the channel's coherence time and assume that the multipath channel has $K=\mathrm{7}$ MPCs with Rayleigh distributed magnitudes. The delays of MPCs are set to $\mathrm{\{3, 5, 10, 16, 22, 28, 33\}}$ ns and their average powers are set to $\mathrm{\{0, -3, -5, -4, -6,-5.5, -7\}}$ dB. The number of iterations allowed for convergence of the SNLS problem (\ref{eq:sub_fit_est1}) is set to 10. To assess the performance of the algorithm, we compute root mean square error (RMSE) of the LOS delay estimate using $10^4$ Monte Carlo trials and compare it to the CRB derived in Section \ref{sc:crb}. The RMSE is defined as $\textrm{RMSE}(\hat{\btau}) := \sqrt{\textrm{MSE}(\hat{\btau})}$, where MSE is given by (\ref{eq:mse}). In the subsequent simulations, some of these parameters are varied. 

\begin{figure}[t!] 
    \centering
    \includegraphics[width=0.85\columnwidth]{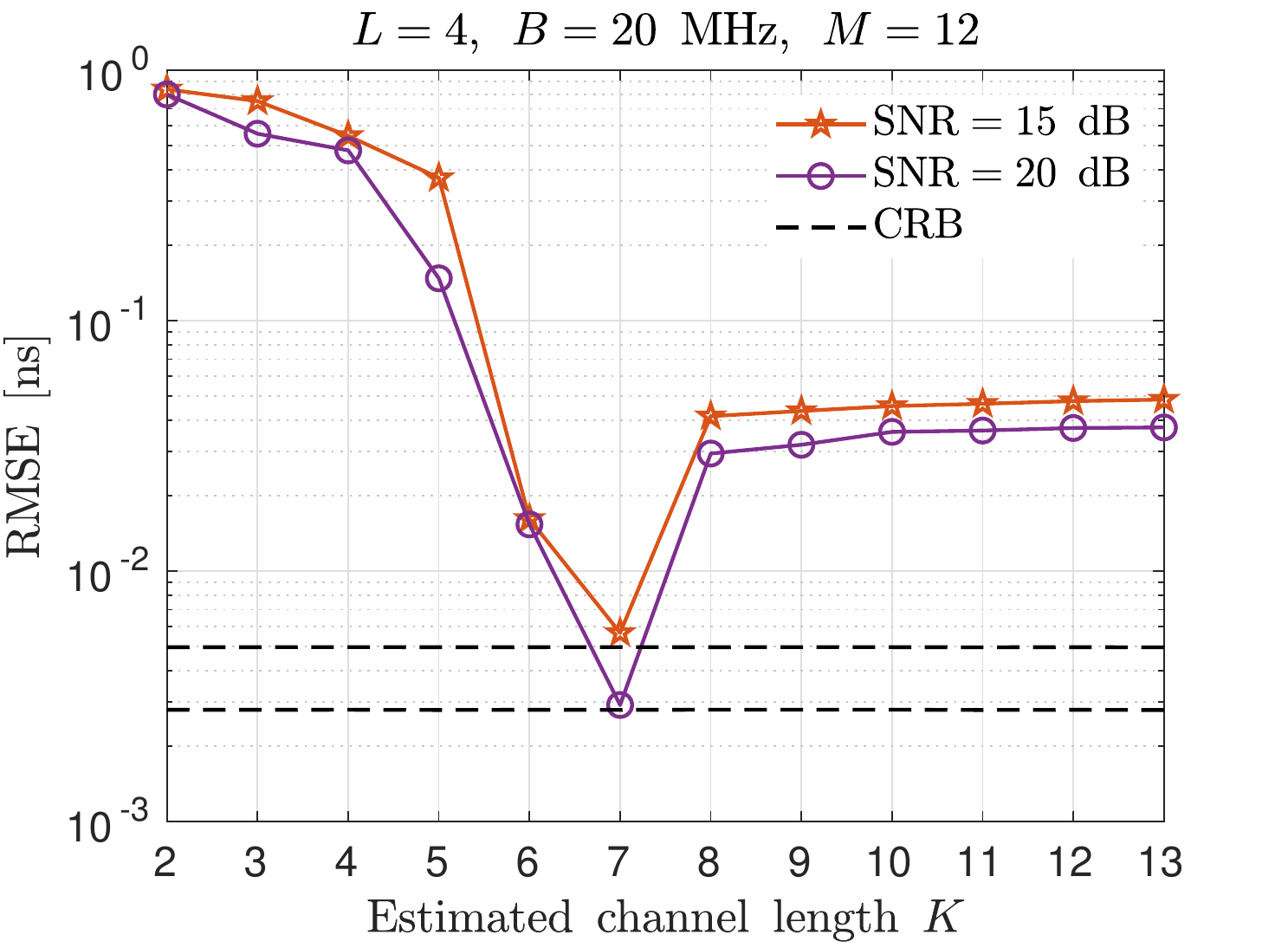}
    \caption{Impact of MPC misdetection on the performance of MBWDE (FB\&NR)}
\label{fig:figure_res3a}
\end{figure}

\begin{figure}[t!] 
    \centering
    \includegraphics[width=0.85\columnwidth]{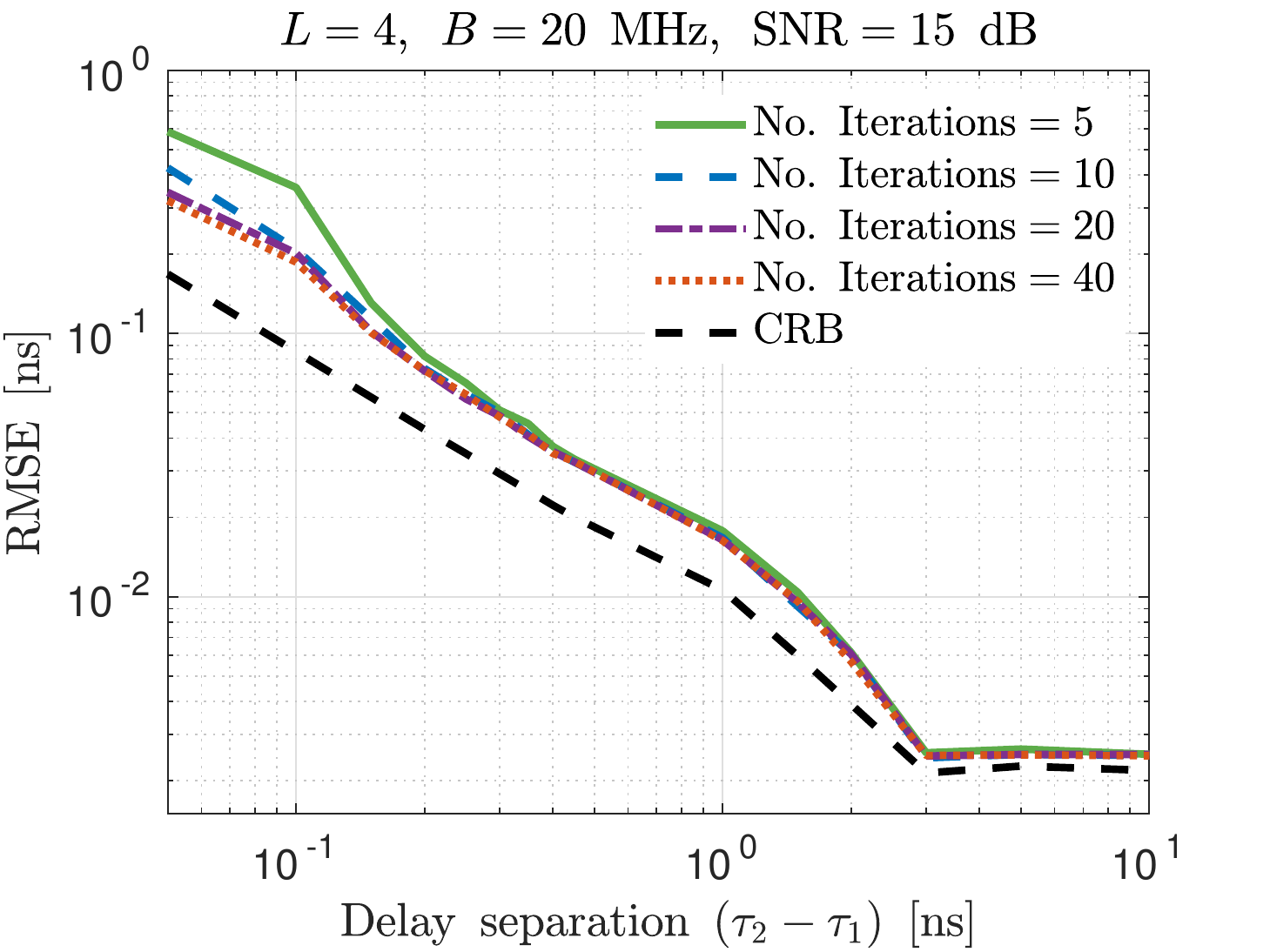}
    \caption{RMSE of estimation for MBWDE(FB\&NR) considering delay separation between the LOS path and the closest MPC.}
\label{fig:fig_res2}
\end{figure}

\subsubsection{Performance of MBWDE}
Fig. \ref{fig:fig_res1} shows the RMSE of delay estimation for different variants of the MBWDE algorithm and its initialization obtained using MR delay estimation as a function of SNR. The unweighted variant of the algorithm is indicated with MBDE, and the variants that include FB averaging and noise reduction or both have extensions (FB), (NR), and (FB\&NR), respectively. All simulation parameters are set as listed previously. It is seen that the MBWDE algorithm asymptotically achieves the CRB as the SNR increases. The results also show that FB averaging and NR techniques provide approximately 2.5 dB of SNR gain. The MBWDE(FB\&NR) variant of the algorithm performs the best, and in the following, we will mostly focus on it.

\subsubsection{Influence of System Parameters $M$, $B$ and $\{f_{c,i}\}_{i=1}^L$}
We first study the scenario where all parameters are set as in the default setup, except that now we vary the number of CSI snapshots. We repeat these simulations for SNR = 5 and 15 dB and compare the performance of MBDE(FB), MBWDE(FB), and MBWDE(FB\&NR). From Fig. \ref{fig:fig_res5}, it is seen that the performances of all algorithms improve when the number of CSI snapshots is increased. However, MBDE(FB) never achieves the CRB and stays biased, even for high SNR. On the other hand, 12 snapshots are enough for the MBWDE(FB) and MBWDE(FB\&NR) algorithms to attain the CRB for high SNR (15 dB), while for low SNR (5 dB), these algorithms attain the bound for 30 snapshots and more. 

Next, we simulate the scenario where the bandwidth of training signals is varying. We set the bandwidth parameter $B$ to $\mathrm{\{20, 40, 80, 160\}}$ MHz. The other parameters are set as in the default simulation setup. Fig.\ \ref{fig:figure_res6a} shows the RMSE of the delay estimation for the  MBWDE(FB\&NR) algorithm. As expected, it is seen that by increasing the bandwidth, the resolution increases. A gain of approximately 10 dB in SNR is achieved when the bandwidth $B$ is doubled.

We have shown in Section \ref{sc:crb} that by increasing the frequency aperture of the CSI measurements, the CRB decreases. Now, we simulate scenarios where the carrier frequencies of the bands are set to the following sets: $\{6, 6.120, 6.320, 6.440\}$, $\{6, 6.160, 6.320, 6.440\}$ and $\{5.960, 6.120, 6.320, 6.480\}$ GHz. Fig.\ \ref{fig:figure_res6b}  shows that the resolution of estimation increases for larger frequency apertures.  However, it is also seen that for low SNR, RMSE increases for a larger aperture. This confirms the results of Section \ref{sc:crb}, and we can conclude that the band selection is a trade-off between resolution and robustness to noise. \TA{Later, in Section \ref{sc:real_bandselection} we will see that real multipath channels are frequency-dependent, which sets a limit on the size of frequency aperture that can be selected without introducing modeling errors in (\ref{eq:multiband}).} 

\begin{figure}[t!] 
    \centering
    \includegraphics[width=0.85\columnwidth]{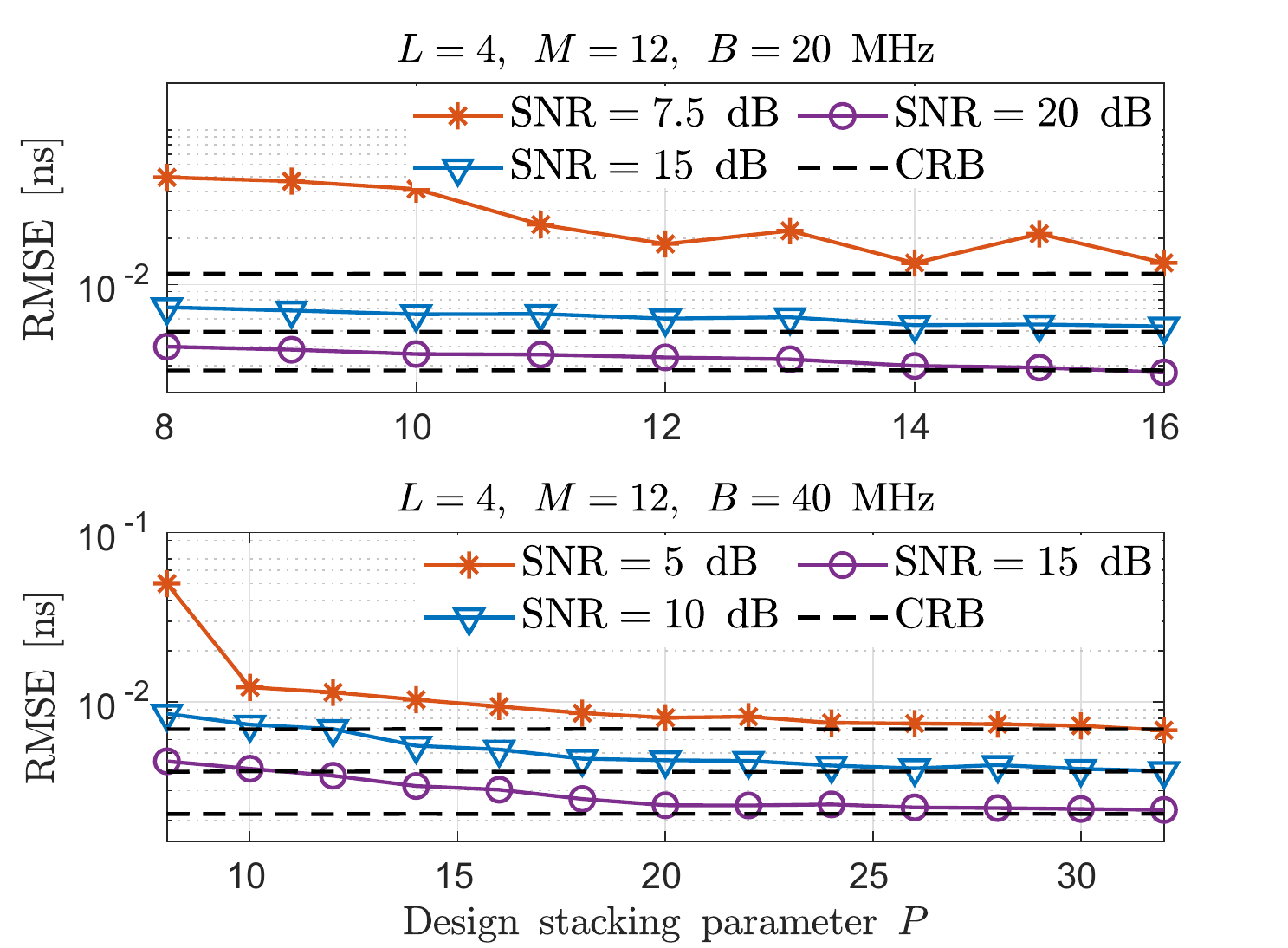}
    \caption{Influence of stacking design parameter $P$ on the RMSE.}
\label{fig:fig_res4}
\end{figure}

\subsubsection{Influence of MPC Detection}
Fig. \ref{fig:figure_res3a} shows the RMSE of the MBWDE algorithm when the number of MPCs in the channel $K$ is wrongly detected. We consider two scenarios where the value of SNR is set to 15 and 20 dB. The true number of MPCs $K$ is 7. It is seen that when $K$ is correctly detected, the algorithm attains the CRB. Its performance sharply deteriorates when $K$ is wrongly detected. The underestimation of $K$ introduces modeling error, and it is more severe compared to overestimation.


\subsubsection{Resolution and Convergence of MBWDE}
\label{sc:resconv}
We assess the resolution of the MBWDE(FB\&NR) algorithm by varying the delay separation between LOS and the closest MPC, i.e., $\Delta \tau_{2,1}= \tau_2-\tau_1$, in the range from 0.01 to 10 ns, while keeping the SNR fixed at 15 dB. We repeat this simulation scenario while setting the number of iterations allowed for convergence of the SNLS problem to $\{5, 10, 20, 40\}$. Fig. \ref{fig:fig_res2} shows the RMSE for this scenario, and it can be seen that the algorithm converges to the CRB for delay separation higher than 2 ns. It is also seen, that for "well-separated" paths ($\Delta \tau_{2,1} \geq 2$ ns), 5 iterations are sufficient for the algorithm (\ref{eq:sub_fit_est1}) to converge.  For critical scenarios, when paths are closely spaced ($\Delta \tau_{2,1} \leq 0.2$ ns), there is a slight improvement when 10 or more iterations are allowed for convergence. However, allowing more than 10 iterations does not result in substantially better performance. \TA{This experiment illustrated the impact of the first MPC on the delay estimation of the LOS path. In \cite{kazaz2020}, we analyzed the impact of other MPCs of delay estimation of the LOS path using MBWDE(FB\&NR)  algorithm. There we used the idea of the first contiguous cluster \cite{shen2010fundamental} and showed that all the MPCs that are within this cluster, i.e., that are separated less than $1/(BL)$ from the LOS path, introduce bias in delay estimation of the LOS path. This bias depends on the relative power of the MPCs compared to the LOS path and their delay separation from it.}

\begin{figure}[t!] 
    \centering
    \includegraphics[width=0.85\columnwidth]{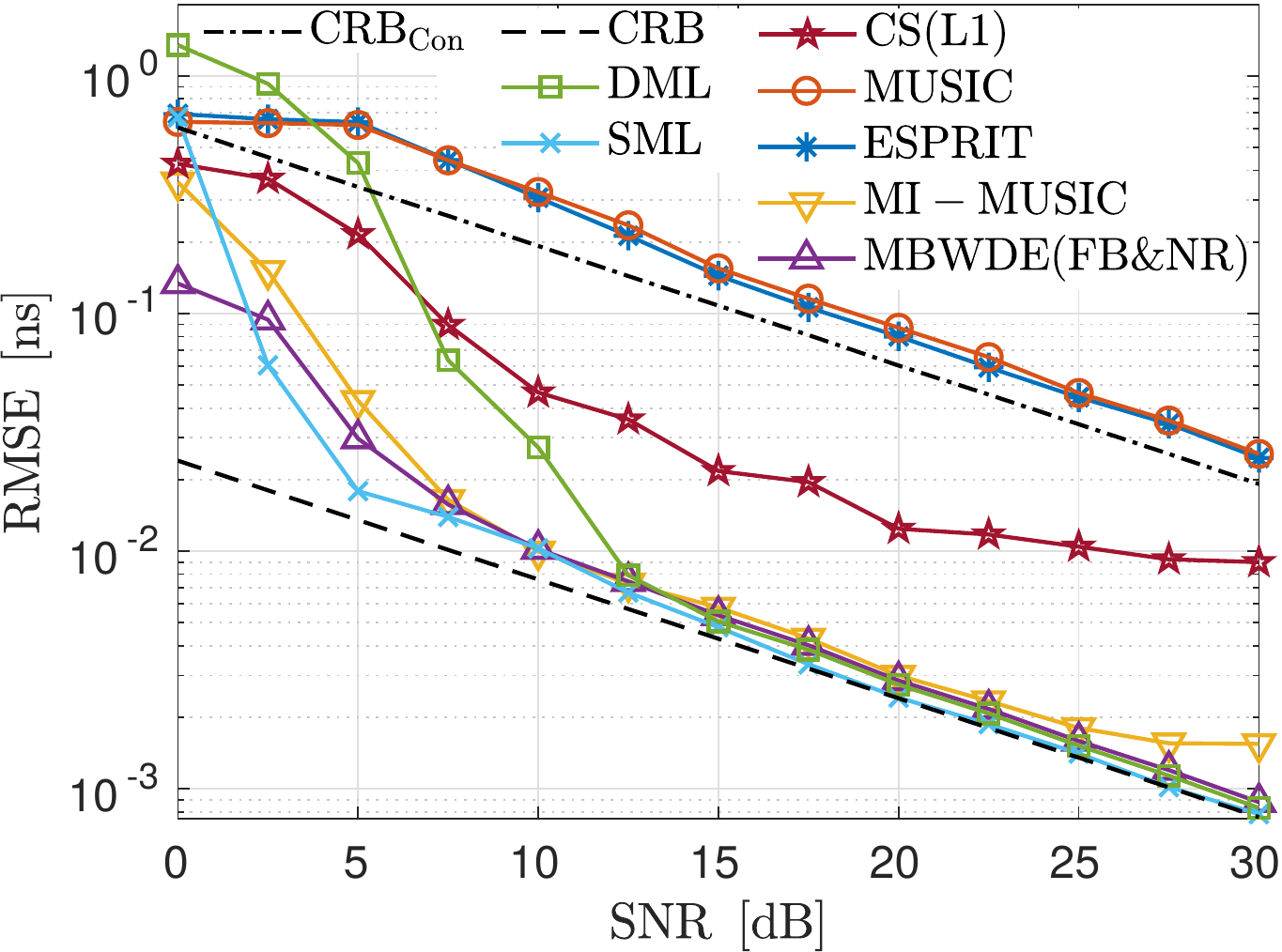}
    \caption{Performance comparison of MBWDE(FB\&NR) with ESPRIT, MUSIC, \TA{DML, SML,} MI-MUSIC and CS(L1).}
\label{fig:fig_res7}
\end{figure}

\subsubsection{Influence of Design Parameter $P$}
In Section \ref{sc:algoritham_outline}, we have introduced the design parameter $P$, which controls the dimensions of the Hankel matrices (\ref{eq:hankel}). We use the default simulation setup to evaluate the influence of parameter $P$ on the RMSE of the algorithm. From Fig. \ref{fig:fig_res4} it is seen that for high SNR, the performance improves when $P$ is increased. This result is intuitive as an increased number of rows in the Hankel matrices increases the frequency aperture. Furthermore, the matrix $\bA'$ (\ref{eq:H_fac}) becomes taller, and the mutual linear independence of its columns increases.

\subsubsection{Comparison to Other Algorithms}
\TA{Finally, we compare MBWDE to DML \cite{dml1} and SML \cite{sml1} methods, algorithms proposed in \cite{tadayon2019decimeter} (MUSIC), \cite{vasisht2016decimeter, khalilsarai2019wifi} (CS(L1)), and DOA estimation algorithms ESPRIT \cite{roy1989esprit}, and MI-MUSIC  \cite{swindlehurst2001exploiting} that are tailored to the problem of delay estimation. We provide CSI with a contiguous bandwidth of $B$ = 80 MHz to MUSIC and ESPRIT.  For all other algorithms, we provide multiband CSI collected in $L$ = 4 bands with $B$ = 20 MHz. The CRB is computed for both contiguous and non-contiguous band allocations. We use delay estimates obtained using MR algorithm \cite{kazaz2019multiresolution} to initialize DML, SML, and MBWDE(FB\&NR). Fig.  \ref{fig:fig_res7} shows that algorithms that utilize contiguous bands have more than a 10 times higher RMSE compared to algorithms that use multiband CSI. The best performance has the SML, which is asymptotically consistent and statistically efficient as $M$ and $B$ tend to infinity. However, it has higher complexity than MBWDE(FB\&NR) as it minimizes a complex multimodal cost function for delays, complex amplitudes, and noise. The performance of MBWDE(FB\&NR) is close to SML. It is also seen, that the consistency and efficiency properties do not hold for the DML in multiple snapshot scenarios, which is shown in Fig. \ref{fig:fig_res7}. The results show that CS(L1) never attains the CRB due to basis mismatch and that MI-MUSIC diverges from it for high SNR ($>$23 dB) where grid mismatch errors dominate noise errors. These errors are caused by the discretization of the delay grid, where we set the grid step to 0.005 ns.  For lower SNR, the performance of MBWDE(FB\&NR) and MI-MUSIC are almost the same. However, MI-MUSIC has a much higher computational complexity due to an exhaustive grid search. }

\begin{figure}[t!] 
    \centering
    \includegraphics[width=0.85\columnwidth]{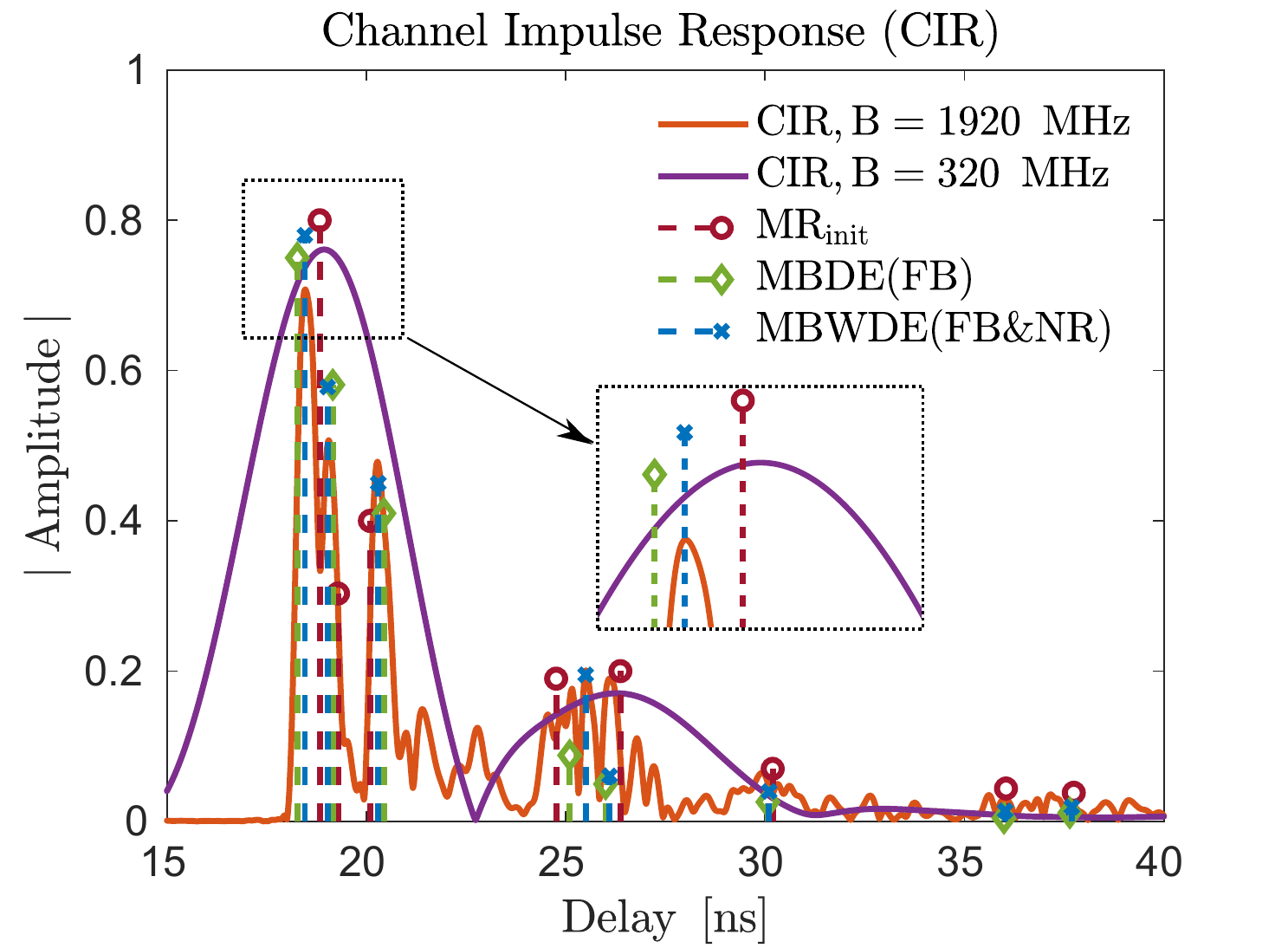}
    \caption{Channel impulse response for varying total bandwidth and estimates of MPCs in a university building.}
\label{fig:fig_exp0}
\end{figure}

\begin{figure} 
    \centering
    \includegraphics[width=0.8\columnwidth]{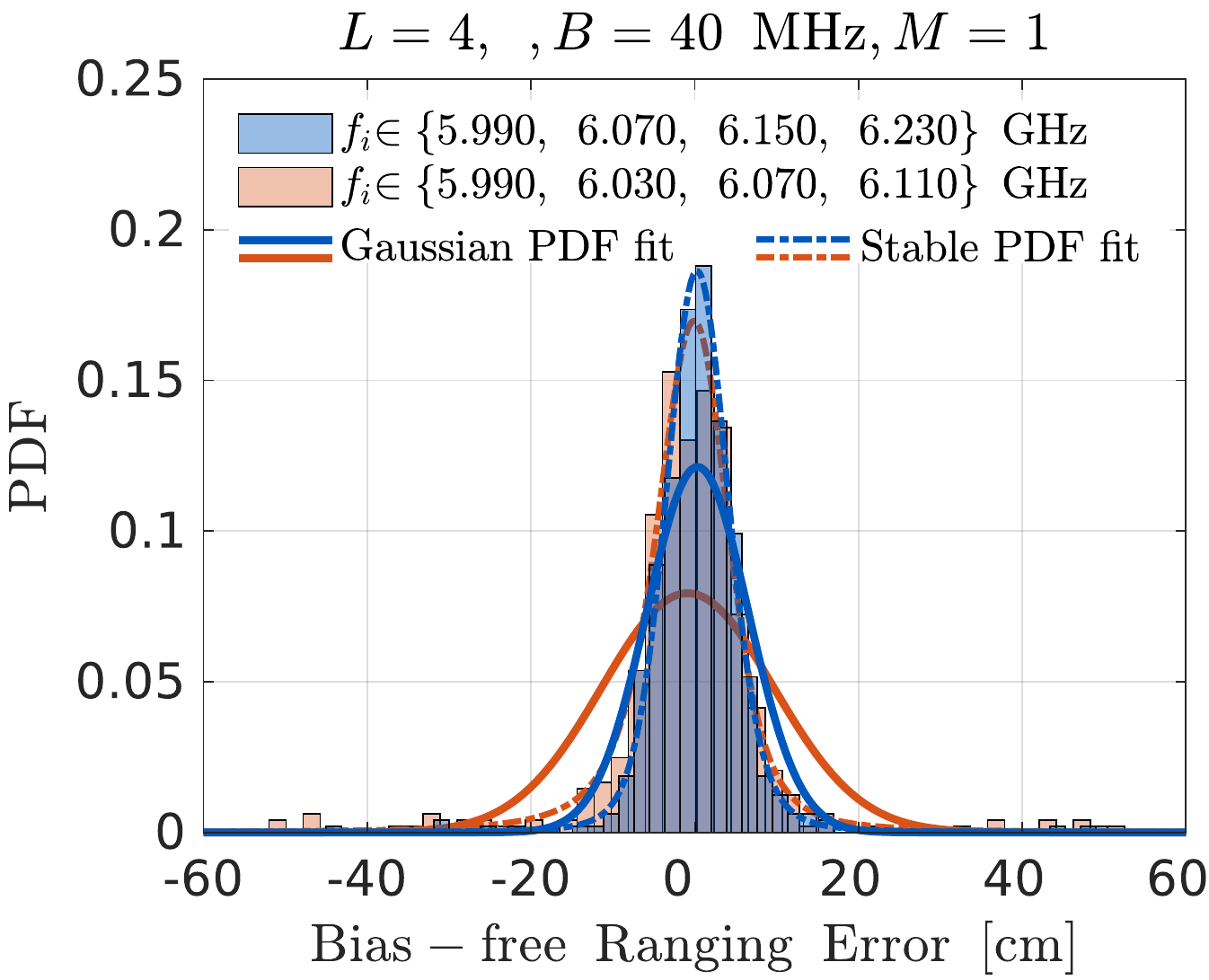}
    \caption{Histogram of ranging errors with respect to band selection fitted with Gaussian and Stable PDFs.}
\label{fig:fig_exp2}
\end{figure}

\section{Real Data Experiments}
\label{sc:real_exp}
In this section, we present experiments with real channel measurements that show the performance of the algorithm in practical scenarios and verify the modeling assumptions made in Section \ref{sc:system_model}. First, we describe the experimental setup used to collect measurements and then present experiments that illustrate the impact of band selection on the RMSE of the range estimates. Later, we use the empirical cumulative distribution function (CDF) of range estimation errors to compare the algorithm with several other methods. Finally, we illustrate the performance of 2-D positioning based on estimated ranges between a mobile node and anchors. 

\TA{We use two indoor CFR datasets collected using a vector network analyzer (VNA) in a hospital \cite{romme2014measurement}, and a university building environment \cite{mint}. The single snapshots of CFR measurement are collected between multiple anchors and a mobile node moving on predefined trajectories in two different indoor environments. The CFRs are measured on a discrete set of equispaced frequencies, which is equivalent to CSI estimation on OFDM subcarrier frequencies using the training signals such as EHT-LTF used in IEEE 802.11be transceivers.}

\TA{When collected with off-the-shelf transceivers, the CSI measurements might get affected by various hardware impairments. A detailed discussion on these effects is provided in \cite{tadayon2019decimeter}. In Section \ref{sc:system_model}, we discussed how to calibrate some of these effects, such as nonideal frequency response of RF chains, phase, and frequency offsets. The CFR measurements that we consider in this section are calibrated up to the effects of antennas. However, the effects of the antennas will introduce an unknown bias in the range estimates. We compute this bias as the mean error of the range estimates compared to the ground truth and eliminate it from the estimates. The other approach would be to directly estimate bias from the range estimates using the known position of the anchors and multidimensional scaling \cite{zhao2019calibration}. The two datasets that we consider are collected using different antennas. Therefore, the calculated biases are different, and they are equal to 5 cm for the hospital and 4.3 cm for the university building environment. The calculated biases stay constant for each of the anchors in a single dataset as they all have the same antenna.}

\begin{table}[]
    \minipage[t]{0.5\textwidth}
        \centering
        \caption{Statistical parameters of ranging errors in a university building for selection of $
        f_{c,i}$ (a) and $B$ (b)}
        \label{tab:tab1}
            \subfloat[][]{ 
                \small
                \begin{tabular}{|c|c|c|}
                \hline
                 \multicolumn{3}{|c|}{$\lvert\hat{d}-d\lvert$}\\
                \hline
                \textbf{Scenario} & \makecell{\textbf{Median} \\ {[}cm{]}} & \makecell{\textbf{Q95} \\ {[}cm{]}} \\
                \hline
                1. & 7.74 & 43.32 \\
                2. & 6.87 & 19.55 \\ \hline
                \end{tabular}
            }
            \hfill
            \subfloat[][]{  
                \small
                \begin{tabular}{|c|c|c|}
                \hline
                 \multicolumn{3}{|c|}{$\lvert\hat{d}-d\lvert$}\\
                \hline
                \textbf{Scenario} & \makecell{\textbf{Median} \\ {[}cm{]}} & \makecell {\textbf{Q95} \\ {[}cm{]}} \\
                \hline
                1. & 7.95 & \multicolumn{1}{c|}{38.37} \\ 
                2. & 6.87 & \multicolumn{1}{c|}{19.55} \\ 
                3. & 4.05 & \multicolumn{1}{c|}{10.84} \\ 
                4. & 0.86 & \multicolumn{1}{c|}{2.89}  \\ \hline
                \end{tabular}
            }
    \endminipage
\end{table}

\begin{table}[]
    \minipage[t]{0.5\textwidth}
        \centering
            \caption{Statistical properties of 2-D positioning error in a indoor hospital environment for several choices of parameter $B$}
             \label{tab:tab2}
            \small
            \begin{tabular}{|c|c|c|c|c|}
            \hline
                & \multicolumn{4}{c|}{ $\left\lVert \hat{\bp}-\bp \right\rVert_2$} \\ \hline
            \textit{B} {[}MHz{]} &
              \begin{tabular}[c]{@{}l@{}}\makecell{\textbf{Mean} \\ {[}cm{]}}\end{tabular} &
              \begin{tabular}[c]{@{}l@{}}\makecell{$\boldsymbol{\sigma_{\rm p}}$ \\ {[}cm{]}}\end{tabular} &
              \begin{tabular}[c]{@{}l@{}}\makecell{\textbf{Q80} \\ {[}cm{]}}\end{tabular} &
              \begin{tabular}[c]{@{}l@{}}\makecell{\textbf{Q95} \\ {[}cm{]}}\end{tabular} \\ \hline
            20  &  26.23      & 17.21       & 30.23 & 47.95       \\ 
            40  &  17.61      & 11.02       & 22.91 & 30.23       \\ 
            80  &  9.97         & 6.28        & 16.87 & 23.57       \\
            160 &  4.82       & 2.72        & 7.06  & 9.29        
            \\ \hline
            \end{tabular}
    \endminipage
\end{table}

\begin{figure*} 
\centering
    \subfloat[]{\includegraphics[trim=0 0 1.2cm 0,clip,width=0.68\columnwidth]{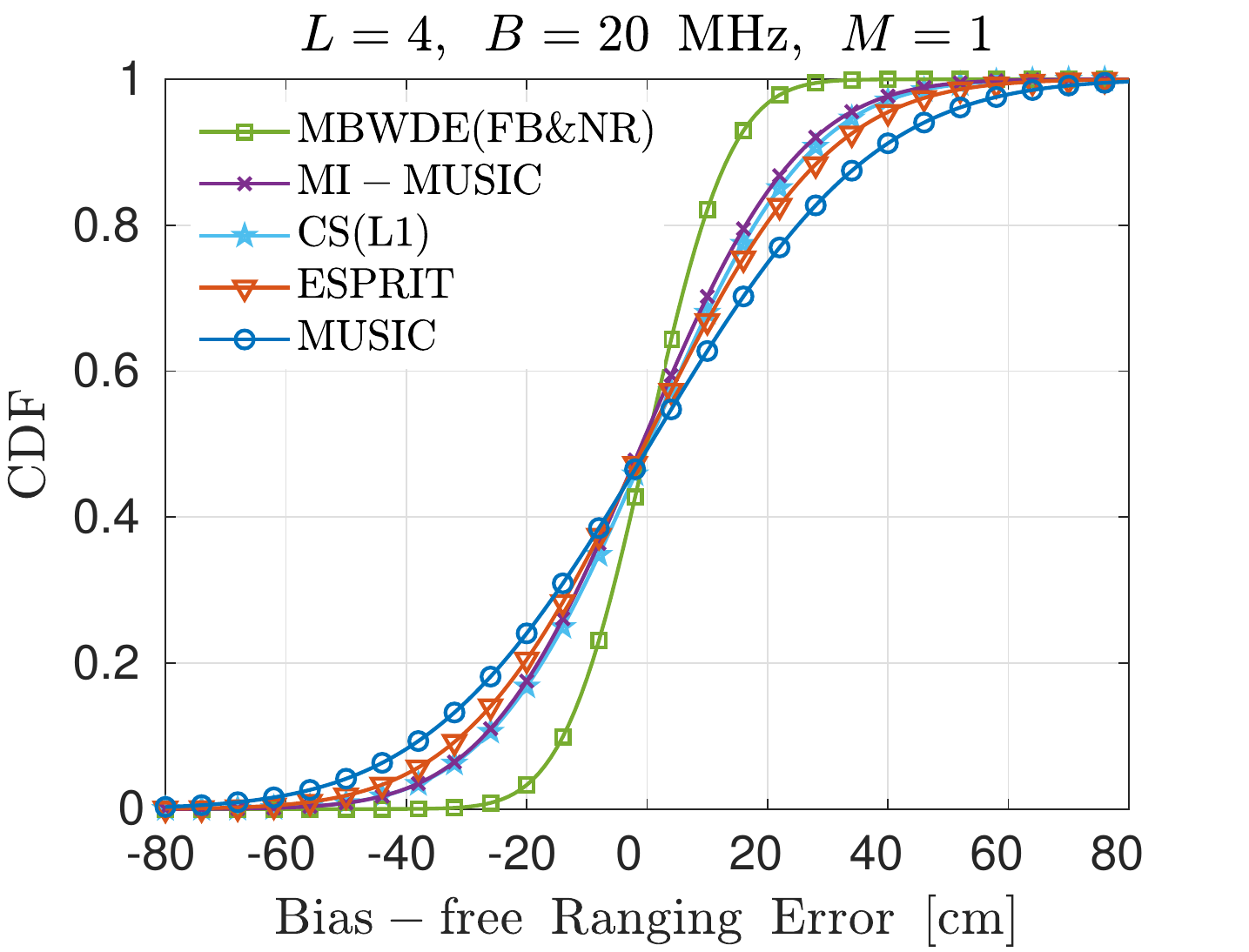}}
    \hfil
    \subfloat[]{\includegraphics[trim=0 0 1.2cm 0,clip,width=0.68\columnwidth]{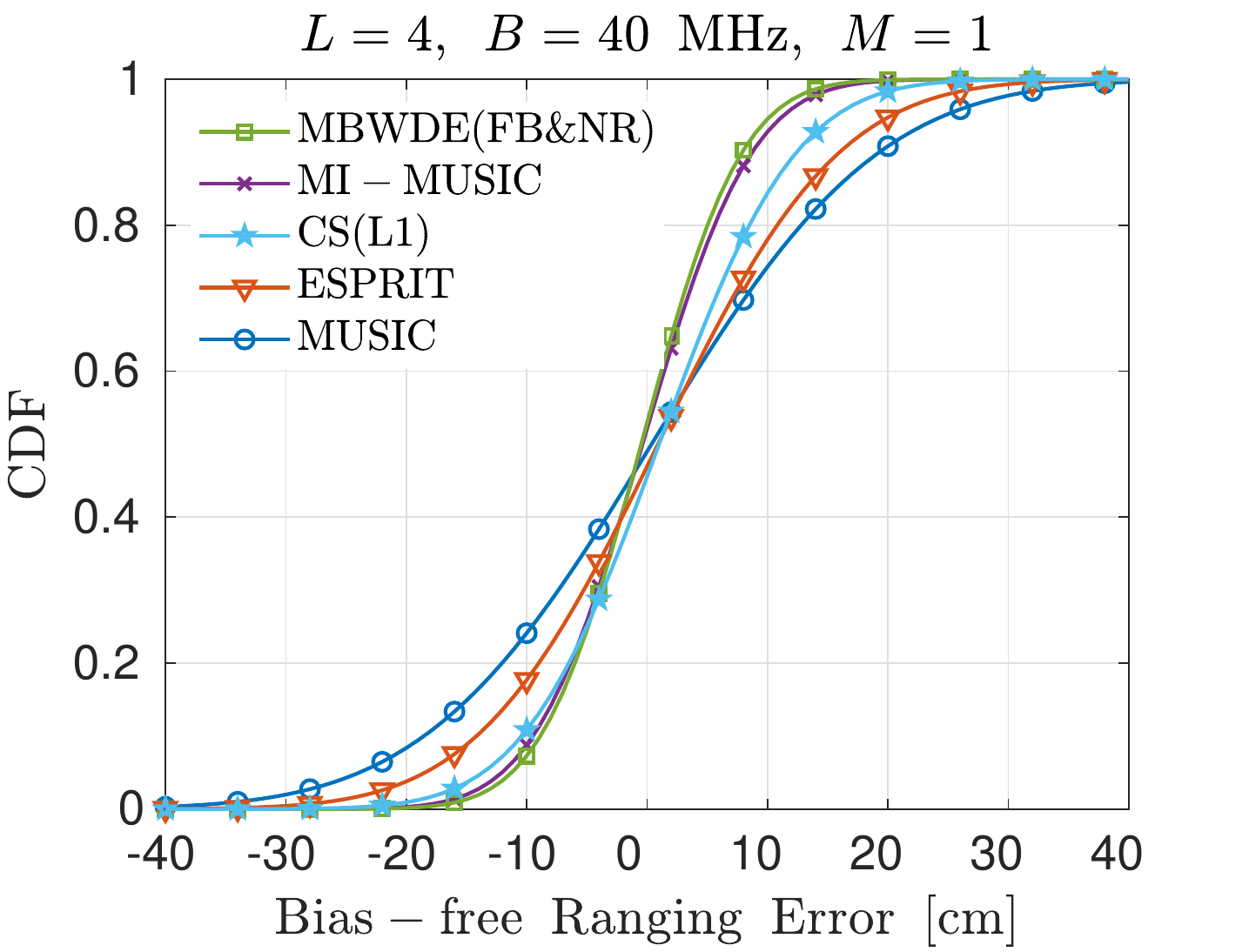}}
    \hfil
    \subfloat[]{\includegraphics[trim=0 0 1.2cm 0,clip,width=0.68\columnwidth]{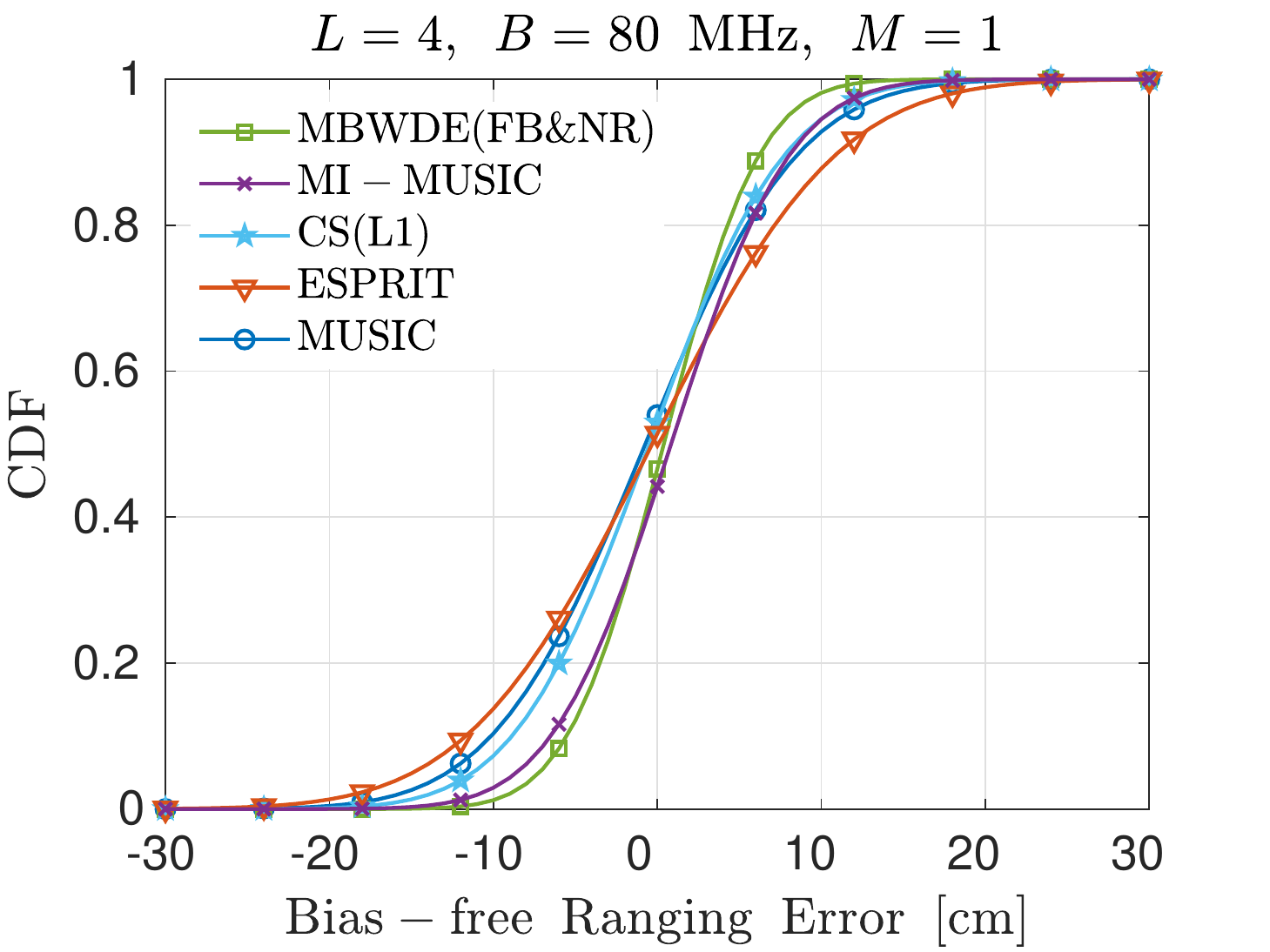}}
    \caption{The CDFs of ranging errors of compared algorithms and choices of parameter $B$ 20 (a), 40 (b), and 80 MHz (c).}
\label{fig:fig_exp1}
\end{figure*}

\subsection{Influence of System Parameters $\{f_{c,i}\}_{i=1}^L$ and $B$}
\label{sc:smi_inf_sysPar}
In this experiment, we use CFR measurements collected between a single anchor and a mobile node in the hallway of a university building, where the mobile node is moving in an area of 1 $\mathrm{m^2}$ \cite{mint}. In total, 484 CFRs are collected for different positions of the mobile node. The CFR is measured over 7501 discrete and equispaced frequencies with the spacing of 1 MHz, starting from 3.1 to 10.6 GHz. This frequency spacing is 12.8 times larger than the subcarrier spacing of 78.125 kHz used in EHT-LTF. The larger number of CFR samples collected using EHT-LTF will slightly improve the RMSE of range estimates with respect to the noise. However, it will not impact resolution as the bandwidth of the measurements is the same. The transmit power of the training signal is set to +15 dBm. We use this experiment to illustrate the effects of the band and bandwidth selection on the RMSE of the range estimates. \TA {We control the bandwidth by varying the number of discrete frequency points (i.e., subcarriers) on which CFR is estimated.}

Fig.\ \ref{fig:fig_exp0} shows the influence of bandwidth selection on the CIR. The CIRs are computed using the CFRs with bandwidths of 1920 and 320 MHz for one of the mobile node positions. \TA{The figure also shows the estimates of MPCs obtained using MR, MBDE, and MBWDE algorithms. The number of bands for the MR algorithm is set to $L$ = 2 and bandwidth to $B$ = 80 MHz. Similarly, for MBDE and MBWDE algorithms, the number of bands is set to $L$ = 4 bands, and bandwidth stays the same as for the MR algorithm.}. The CIR for the bandwidth of $B$ = 320 MHz shows that the LOS path and first two MPCs are not resolved, which would result in biased delay estimation with traditional methods. On the other hand, it is seen that the MBWDE algorithm almost perfectly estimates the delay of the LOS path for the same total bandwidth.

\begin{figure} 
    \centering
    \includegraphics[trim=0 0 1.2cm 0,clip,width=0.8\columnwidth]{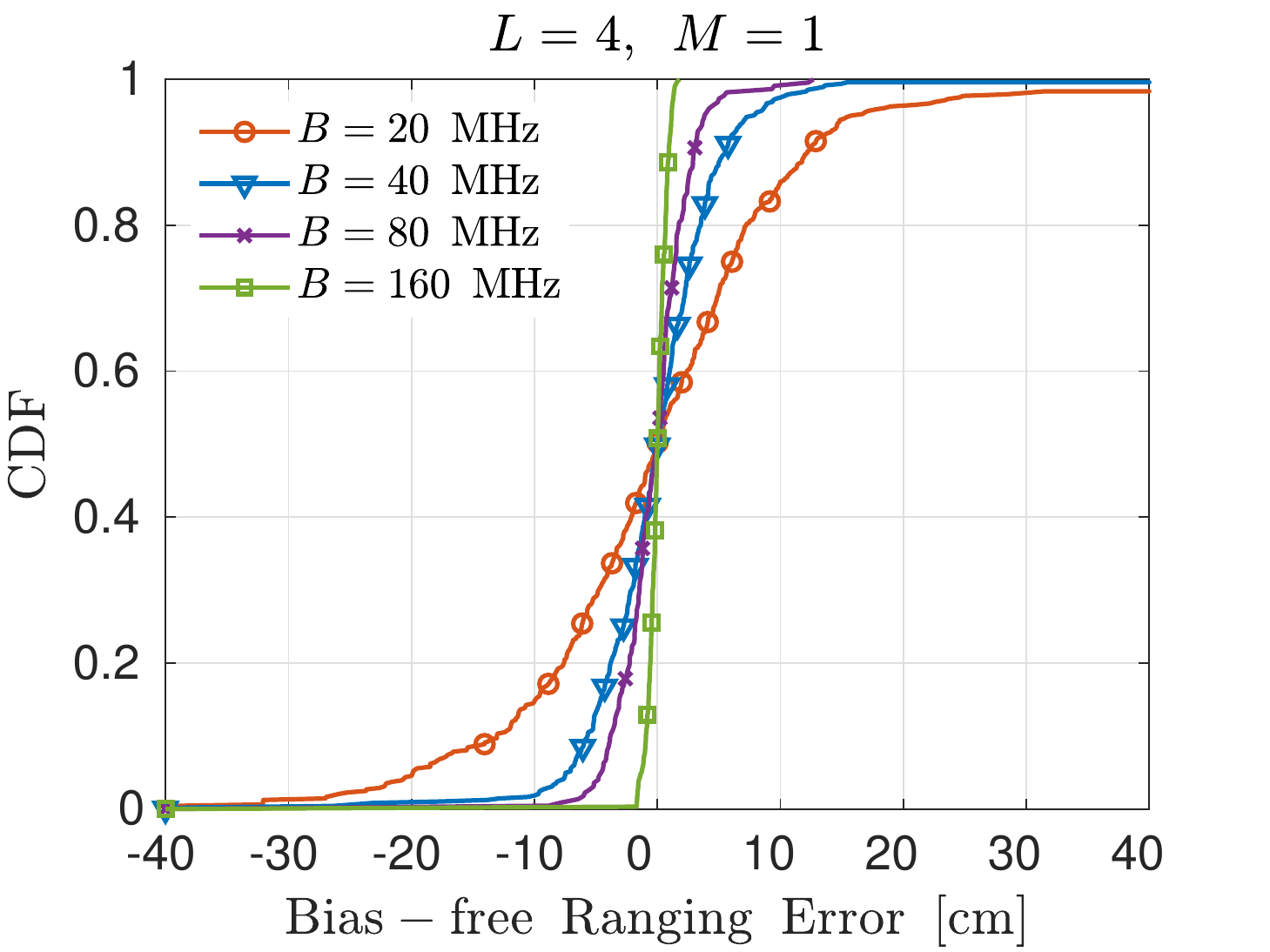}
    \caption{Empirical CDFs of ranging errors with respect to parameter $B$.}
\label{fig:fig_exp21}
\end{figure}

\subsubsection{Influence of Band Selection}
\label{sc:real_bandselection}
We first analyze the distribution of ranging errors with respect to band selection. We consider scenarios where a single snapshot, i.e., $M$ = 1, of CSI measurements is collected in $L=\mathrm{4}$ bands, each with a bandwidth of $B=\mathrm{40}$ MHz. In the first scenario, we collect CSI by taking samples of CFR in the following bands: $f_i \in \mathrm{\{5.990, 6.070, 6.150, 6.230\}}$ GHz. In the second scenario, we lower the total frequency aperture and collect CSI in the following consecutive bands $f_i \in \mathrm{\{5.990, 6.030, 6.070, 6.110\}}$ GHz. The selected frequencies correspond to the IEEE 802.11be channels in the 6 GHz band (cf. Fig. \ref{fig:fig_ilu3}). We estimate the ranges between mobile node and anchor for 484 different locations and compute the ranging errors by comparing estimates with the ground truth. We then estimate the bias as a median value of estimated ranges and compensate for it. Fig. \ref{fig:fig_exp2} shows histograms of bias-free ranging errors normalized to probability. The histograms are fitted to Gaussian and L\'{e}vy alpha-stable distributions. It is seen that due to the small number of outliers with high ranging error, the Gaussian distribution does not perfectly fit the histograms. The alpha-sable distribution is more general compared to Gaussian, and its stability parameter $\alpha$ is tuned to introduce heavy tails in the PDF that better fit outliers \cite{liang2013survey}. The estimated parameters $\alpha$ for the first and second scenarios are 1.72 and 1.45, respectively. However, for these values of $\alpha$, the common properties of distributions such as mean and standard deviation are undefined. Therefore, we use the median value to express bias and 95\%-quantile (Q95) to express the accuracy of the estimates. The Q95 is defined as the segment around the median, which contains 95\% of the estimates. To calculate Q95, we subtract the 2.5$\mathrm{^{th}}$ percentile from the 97.5$\mathrm{^{th}}$ percentile. The median and Q95 for these scenarios are given in Table \ref{tab:tab1} (a). As expected from the results shown in Section \ref{sc:smi_inf_sysPar} for larger frequency aperture, the resolution of the estimates increases, and Q95 is 19.55 cm. This is approximately two times lower compared to Q95 of 43.32 cm obtained using a smaller frequency aperture.

However, the experiments have also shown that selecting a too large frequency aperture can lead to degradation of delay estimation. This is caused by the frequency dependency of RF scattering, which introduces errors in the model  (\ref{eq:multiband}). The same effect occurs in channel extrapolation for FDD massive MIMO systems \cite{rottenberg2020performance}, where the goal is to infer CSI at the downlink band based on CSI estimates from the uplink band. The frequency dependency is hard to model as it depends on materials that are producing an RF scattering scene \cite{haneda2012modeling}. However, these modeling errors are not critical if the frequency aperture is less than 10\% of the carrier frequency \cite{molisch2009ultra}. \TA{We do not optimize the band selection in this work with respect to the trade-off between delay resolution and modeling errors, and this remains an open question for future research. However, we avoid modeling errors in the experiments by estimating ranges from the CSI measurements collected in the bands that create a frequency aperture smaller than 600 MHz.}

\subsubsection{Influence of Bandwidth Selection}
\label{sc:infbwsel}
Next, we consider four scenarios where bandwidth $B$ of $L$=4 bands is varied, and it takes values $\mathrm{\{20, 40, 80, 160\}}$ MHz where their central frequencies are set to $\mathrm{\{5.98, 6.06, 6.14, 6.22\}}$, $\mathrm{\{5.95, 6.03, 6.15, 6.23\}}$, $\mathrm{\{5.97, 6.13, 6.21, 6.37\}}$ and $\mathrm{\{6.01, 6.21, 6.37, 6.57\}}$ GHz, respectively. We repeat the same procedure as previously to compute median and Q95, and the results are shown in Table \ref{tab:tab1}(b). As expected, it is seen that Q95 decreases when the bandwidth is increased, where the gain in accuracy is proportional to the increase in bandwidth. To illustrate the distribution of ranging errors, we plot the empirical CDFs for these scenarios in Fig.\ \ref{fig:fig_exp21}. It is seen that in 80\% of the cases, the absolute ranging error is smaller than approximately 16, 8,  4, and 1 cm for $B \in \{20, 40, 80, 160\}$ MHz, respectively.

\begin{figure} 
    \centering
    \includegraphics[width=1\columnwidth]{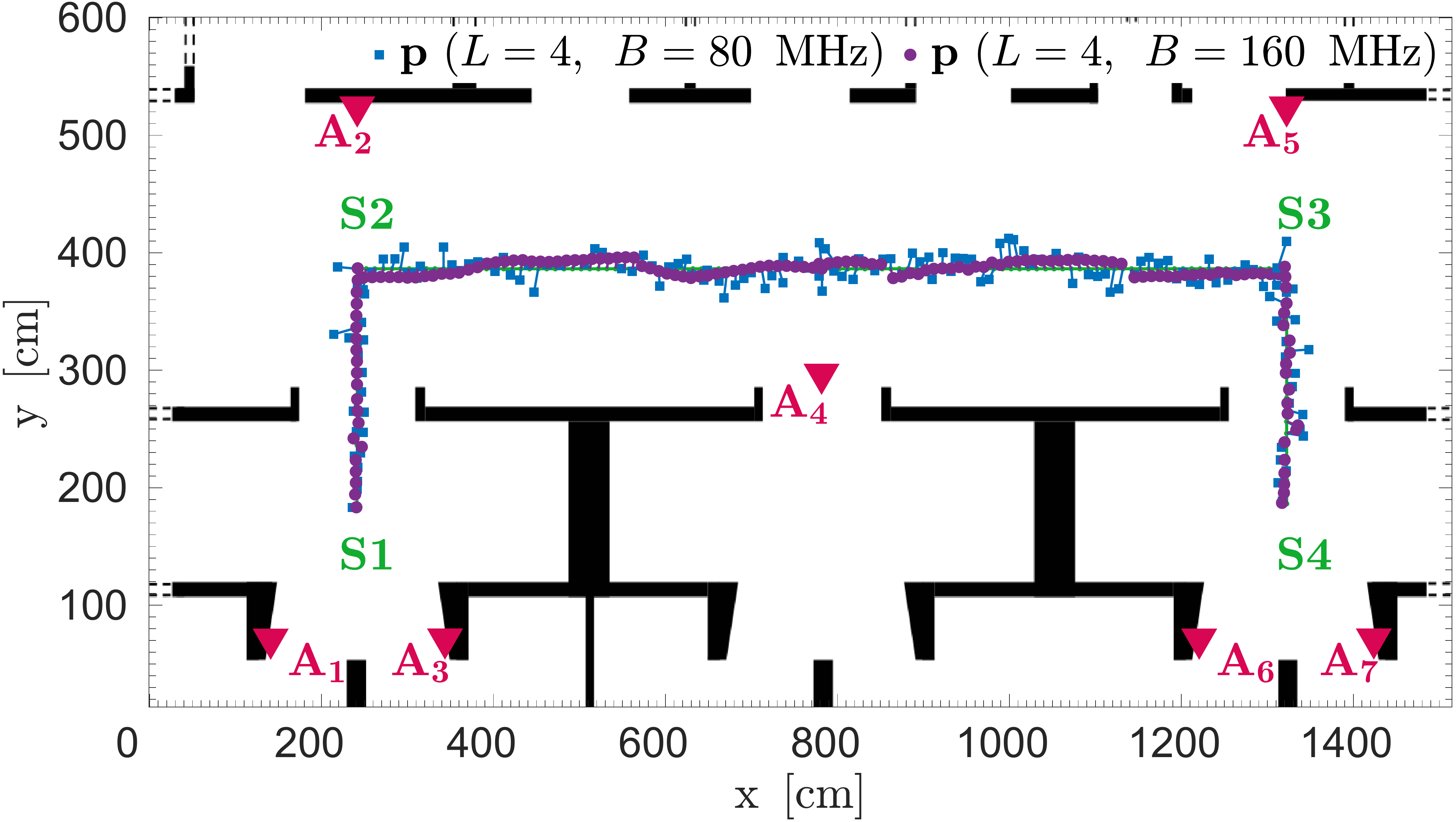}
    \caption{Anchors and trajectory of the mobile node with 2-D position estimates in a hospital building for $B \in \mathrm{\{80, 160\} MHz}$.}
\label{fig:fig_exp6}
\end{figure}

\subsection{Performance of Positioning}
To illustrate the performance of 2-D positioning based on ranges estimated using the MBWDE algorithm, we use CFR measurements collected in an indoor hospital environment. These measurements are collected between 7 anchors and the mobile node for 150 points on a trajectory shown in Fig.\  \ref{fig:fig_exp6}. The CFR is measured over a set of 4096 equispaced frequency points starting from 5 to 10 GHz with a link budget of 110 dB \cite{romme2014measurement}. This is equivalent to a subcarrier spacing of 1.22 MHz, which is 15.6 times higher than the IEEE 802.11be configuration. The same conclusions related to resolution and noise performance as for the previous experiments hold. \TA{To estimate the number $K$ in these experiments, we use the MDL criteria described in Section \ref{sc:algorithm}. The estimated $K$ takes values between 12 and 21 for the trajectory shown in Fig. \ref{fig:fig_exp6}.}

\subsubsection{Comparison to Other Algorithms} 
To compare MBWDE(FB\&NR) algorithm against other algorithms, we consider three scenarios where we vary the bandwidth $B \in \mathrm{\{20, 40, 80\}}$ MHz. The carrier frequencies of the bands for MBWDE(FB\&NR), MI-MUSIC, and CS(L1), are selected as in Section \ref{sc:infbwsel}, while for MUSIC and ESPRIT, a single band with the same total bandwidth is selected starting from 5.925 GHz. The delay grid step is set to 0.15 ns (0.5 cm) for MI-MUSIC, CS(L1), and MUSIC.

We use the previously mentioned algorithms to estimate the ranges between anchor \textit{A2} and the mobile node moving on segment \textit{S1S2S3} on the trajectory (cf. Fig \ref{fig:fig_exp6}), where the segment \textit{S3S4} is omitted due to the presence of NLOS propagation. We compute the ranging error and empirical CDFs in the same way as in previous scenarios. These empirical CDFs are fitted with a Gaussian CDF and shown in  Fig. \ref{fig:fig_exp1}. It is seen that in all scenarios, MBWDE(FB\&NR) has the best performance, where the performance gain is highest for the case when the bandwidth is smallest, i.e., $B$= 20 MHz. In scenarios where $B \in \mathrm{\{40, 80\}}$, the performance of MI-MUSIC and MBWDE(FB\&NR) are almost identical. MUSIC and ESPRIT perform worst for all the scenarios compared to algorithms that use multiband CSI due to the smaller frequency aperture.

\subsubsection{Influence of Ranging on 2-D Positioning}
Finally, we illustrate the performance of 2-D positioning by using range estimates of the MBWDE (FB\&NR) algorithm. We define the mobile node position as $\bp = [x, y]^T$, where $x$ and $y$ are the node's coordinates. The mobile node positions are estimated using a LS algorithm from the ranges estimated between the mobile node and three anchors (cf. Fig. \ref{fig:fig_exp6}). In particular, for the segments \textit{S1S2}, \textit{S2S3}, and \textit{S3S4} the ranges are estimated between mobile node and anchors \{\textit{A1, A2, A3}\}, \{\textit{A2, A4, A5}\} and \{\textit{A5, A6, A7}\}, respectively. \TA{We select anchors based on the floor map shown in Fig. \ref{fig:fig_exp6} to avoid NLOS propagation and outliers in the 2-D positioning. However, when a floor map is not available, this could be done directly from the measurements as shown in \cite{xiao2014non}.}

We estimate positions for four different scenarios with varying bandwidth $B \in \mathrm{\{20, 40, 80, 160\}}$ MHz. Fig. \ref{fig:fig_exp6} shows the position estimates for the scenarios where $B \in \mathrm{\{80, 160\}}$ MHz. It can be seen that for $B = \mathrm{160}$ MHz the position is almost perfectly estimated. To quantify the performance of position estimation, we compute the estimation error as $\textrm{RMSE}(\hat{\bp}) = \left\lVert \hat{\bp}-\bp \right\rVert_2$, where $\bp$ is the ground truth and $\hat{\bp}$ is the estimate. We compute statistical properties of the errors such as mean, standard deviation $\sigma_{\rm p}$, 80\%-quantile (Q80), and Q95. The Q80 is computed by subtracting the 10$\mathrm{^{th}}$ percentile from the 90$\mathrm{^{th}}$ percentile. These properties are given in Table \ref{tab:tab2}. As expected, $\textrm{RMSE}(\hat{\bp})$  decreases with increasing bandwidth $B$. It is seen that with a single snapshot of CSI with a total bandwidth of 320 MHz and using 3 anchor nodes, it is possible to achieve an average positioning error below 24 cm in 95\% of the cases.

\section{Conclusions}
\label{sc:conclusions}
In this paper, we considered high-resolution delay estimation for range-based localization using multiband CSI measurements. We derived a data model for multiband CSI and showed that it has multiple shift-invariance structure. We designed the MBWDE algorithm that exploits this structure to estimate delay parameters. We presented several data extension and preprocessing techniques that further improve the performance of the MBWDE. To assess the performance of the algorithms, we derived the CRB on the RMSE of delay estimates considering a multiband CSI model. We used parameters of the emerging IEEE 802.11be standard to define simulation scenarios that illustrate the performance of MBWDE. The results of simulations showed that MBWDE almost attains the CRB when MPCs present in the channel are resolved and outperforms other multiband estimation algorithms such as CS(L1) and MI-MUSIC. 

To verify the modeling assumptions, we use the MBWDE algorithm to perform ranging and positioning using real indoor multipath channel measurements. These results revealed that the MBWDE algorithm improves the trade-off between delay resolution and bandwidth of the training signals used for CSI estimation. The experiments with real channel measurements also showed that when the total frequency aperture of multiband CSI is increased to more than $\sim$10\% of the carrier frequency, the frequency dependency effects of multipath propagation cause modeling errors that degrade the performance of estimation. Therefore, in the future, it would be of practical interest to model frequency dependency effects of multipath propagation and take these effects into account in MBWDE.


%



\ifCLASSOPTIONcaptionsoff
  \newpage
\fi

\bibliographystyle{IEEEtran}
\bibliography{main}

\end{document}